\documentclass[11pt,journal,onecolumn,letterpaper]{IEEEtran}
\usepackage[letterpaper,margin=1in]{geometry}

\usepackage{amsmath,amssymb,amsfonts,amsthm,mathtools,dsfont}
\usepackage{booktabs,array,tabularx}
\usepackage{graphicx}
\usepackage{cite}
\usepackage{microtype}
\usepackage{needspace}
\usepackage{placeins}
\usepackage[hidelinks]{hyperref}
\usepackage{xcolor}

\allowdisplaybreaks
\theoremstyle{plain}
\newtheorem{theorem}{Theorem}
\newtheorem{lemma}{Lemma}

\theoremstyle{definition}
\newtheorem{definition}{Definition}

\theoremstyle{remark}
\newtheorem{remark}{Remark}

\newcommand{\E}{\mathbb E}
\newcommand{\Prob}{\mathbb P}
\newcommand{\supp}{\operatorname{supp}}
\newcommand{\chom}{C_{\it{hom}}}
\newcommand{\chet}{C_{\it{het}}}

\title{Optimal Communication for Distributed Sparse Histogram Aggregation}
\author{Fangwei Ye, Yijun 
Fan\thanks{
F.~Ye is with the College of Computer Science and Technology, Nanjing University of Aeronautics and Astronautics, Nanjing 211106, China (email: fangweiye@nuaa.edu.cn).

Y.~Fan is with the Department of Information Engineering, The Chinese University of Hong Kong, Hong Kong SAR, China (email: yijunfan@ie.cuhk.edu.hk). 
}
}
\date{}
\begin{document}
\maketitle

\begin{abstract}

Histogram estimation is a fundamental primitive in distributed data analysis, with applications in counting queries and federated learning. In many settings, the universe of categories is enormous, yet each user’s data involves only a small subset, yielding sparse local histograms. This raises a natural question:  To what extent can local sparsity reduce the communication needed to estimate the global histogram? 

We answer this question by characterizing the communication complexity of distributed sparse histogram estimation under prescribed \(\ell_1\) distortion and failure probability constraints.
Our upper bound is achieved by a randomized covering scheme that optimizes how accurately each user represents its local histogram to reduce communication. By exploiting the distortion budget directly rather than imposing unbiasedness, the scheme achieves lower communication costs than standard unbiased sampling and hashing constructions.
We establish the scheme’s first-order optimality through a lower bound based on local support identification. The key step is to relate 
communication cost directly to the number of active coordinates revealed by each user's message, rather than to bound information about the aggregate via distance-based Fano's inequalities.
This characterization further quantifies how the communication advantage of heterogeneous over homogeneous encoders varies across parameter regimes.

\end{abstract}


\newpage

\section{Introduction}
\label{sec:introduction}
 
Histogram estimation asks how often each item appears in a large universe in a collection of data. It supports stream summaries, frequent-item identification, and counting queries~\cite{esfandiari2024coordinator,feng2025heavyhitters,bhattacharyya2018heavyhitters}. When the data are distributed, a server obtains these frequencies through local summaries.
Many attempts have been made to understand the accuracy and communication complexity of the distributed histogram aggregation, such as mergeable sketches~\cite{karnin2016quantiles}, communication-efficient mean estimation~\cite{suresh2017mean,braverman2016distributed}, and federated learning~\cite{mcmahan2017federated}.
A problem of particular interest in practice is that with sparse local histograms, the domain of possible categories is usually very large, while each user's data involve only a small fraction of these categories.
We are interested in 
asking how many bits an accurate aggregate requires.

We consider a distributed system consisting of $n\geq 1$ users and a server. User $i$ holds a $k$-sparse categorical vector
\[
 x_i\in\{0,1\}^d,\qquad \|x_i\|_0\leq k,
 \qquad i=1,\ldots,n,
\]
where $d\geq2$ and $1\leq k\leq d$. The target for the server is to estimate empirical histogram
\begin{equation}
 u=\frac1n\sum_{i=1}^n x_i.
 \label{eq:target}
\end{equation}
The problem is closely related to communication-efficient distributed mean estimation, where sparsity, correlations, and quantization are used to reduce communication~\cite{suresh2017mean,huang2019optimal,jhunjhunwala2021spatial,vargaftik2022eden,suresh2022correlated}. Another related problem is heavy-hitter identification in data streaming, where compact summaries identify items with large frequencies~\cite{bhattacharyya2018heavyhitters}\cite{feng2025heavyhitters}. 
Sparse aggregation through the lens of lossy compression~\cite{fan-cover} and variants of the problem with additional security and privacy constraints~\cite{chen2023security} have also been studied.
Here we focus on characterization of the 
communication complexity for a required accuracy.

The accuracy requirement is defined as follows.
\begin{definition}
For $0\leq D<k$ and $0\leq\delta<1$, a protocol is \emph{$(D,\delta)$-accurate} if
\begin{equation}
 \inf_{\substack{x_1,\ldots,x_n\in\{0,1\}^d\\
                    \|x_i\|_0\leq k\ \forall i}}
 \Prob\left(\|\hat u-u\|_1\leq D\right) \geq 1-\delta.
 \label{eq:accuracy}
\end{equation}
\end{definition}
Our goal is to determine how much communication complexity approximation can save. The guarantee is minimax over arbitrary inputs, including all patterns of support overlap.

Communication complexity here is measured by the expected total number of transmitted bits, allowing random message lengths as in variable length compression with errors~\cite{kostina2015variable} and empirical mean estimation~\cite{konecny2018mean}. Let $R$ be public randomness independent of the inputs and $M_i\in\{0,1\}^*$ be user $i$'s message. A protocol has cost
\begin{equation}
 \sup_{\substack{x_1,\ldots,x_n\in\{0,1\}^d\\
                    \|x_i\|_0\leq k\ \forall i}}
 \E\!\left[\sum_{i=1}^n|M_i|\right].
 \label{eq:cost}
\end{equation}
Both probability and expectation are over protocol randomness. Public randomness is considered free~\cite{acharya2020inference}, i.e., all users and the server have access to public randomness $R$, independent of the inputs, without communication cost.

In distributed estimation, two commonly considered scenarios allow users to employ different encoding rules or require all users to share a common rule~\cite{acharya2019communication}. Here, we study both cases. In the \emph{heterogeneous} model, user $i$ may use its own conditional channel:
\begin{equation}
 \Prob(M_1=m_1,\ldots,M_n=m_n\mid x_1,\ldots,x_n,R=r)
 =\prod_{i=1}^n Q_{i,r}(m_i\mid x_i).
 \label{eq:heterogeneity}
\end{equation}
In the \emph{homogeneous} model, all users use the same channel after fixing the public randomness:
\begin{equation}
 \Prob(M_1=m_1,\ldots,M_n=m_n\mid x_1,\ldots,x_n,R=r)
 =\prod_{i=1}^n Q_r(m_i\mid x_i).
 \label{eq:homogeneity}
\end{equation}
The reason of distinguishing homogeneous and heterogeneous models is because the gap between them has been observed in related problems~\cite{acharya2019communication}.
Let $\chom(D,\delta)$ and $\chet(D,\delta)$ be the infima of \eqref{eq:cost} over accurate homogeneous and heterogeneous protocols, suppressing $n,k,d$. Class inclusion gives
\begin{equation}
 \chet(D,\delta)\leq\chom(D,\delta).
 \label{eq:class-inclusion}
\end{equation}
An interesting finding of our results is that it requires more communication to achieve the same accuracy when using the same encoding rule for all users. 

Throughout this paper, $\log$ are in base $2$ and $\ln$ are in the natural base $e$. Since zero communication suffices at $D\geq k$ or $\delta=1$, we consider the non-trivial cases with positive communication cost needed in the sequel.

\section{Overview}
\label{sec:overview}

\subsection{Prior Works and Existing Results}
\label{sec:related-work}

Previous approaches reduce communication complexity by sampling data or encoding local supports. Coordinate and client sampling give unbiased estimates~\cite{suresh2017mean}\cite{konecny2018mean}, hashing produces compact summaries for histogram estimation~\cite{braverman2016countsketch}\cite{liu2020succinctfilters}, and algebraic syndromes recover sparse supports exactly~\cite{belazzougui2024reconciliation}. Lossless compression can further reduce the size of these representations~\cite{huang2019optimal}\cite{mitzenmacher2002compressed}. We compare these constructions under the same empirical target, expected communication cost, and $(D,\delta)$-accuracy requirement.

\emph{Unbiased coordinate sampling} reports each active coordinate independently with probability $p$ and reconstructs a reported entry as $1/p$~\cite{konecny2018mean}. \emph{Client sampling} selects each user independently with probability $p$. Selected users report their entire supports, and the server applies the same rescaling~\cite{suresh2017mean}. Both protocols cost at most $nkp\lceil\log d\rceil$ expected bits. Their tail guarantees can differ since coordinate sampling keeps each active coordinate independently, whereas client sampling keeps or omits all active coordinates of a user together. Thus, under $(D,\delta)$-accuracy guarantee, their achievable communication complexity are of the same order but not exactly identical.

Collision-based methods instead summarize all active coordinates. \emph{Count-Min Sketch} hashes them into nonnegative counters and queries the minimum across rows~\cite{cormode2005countmin}. Thresholding the answer at one gives no false negatives on binary inputs, and a coordinate is absent only if it collides with an active coordinate in every row. \emph{Compressed Bloom filters} record whether each bucket contains an active coordinate and compress the resulting bit arrays~\cite{bloom1970space}\cite{mitzenmacher2002compressed}. Both use smaller communication cost by allowing more zero coordinates to be incorrectly reported as nonzero.

\emph{BCH support recovery} uses algebraic code and removes the distortion in these local summaries. An algebraic syndrome of at most $k\lceil\log(d+1)\rceil$ bits identifies every support of size at most $k$~\cite[Theorem 16]{belazzougui2024reconciliation}. Decoding each user and averaging gives exact aggregation independent of $D$. When failure probability $\delta$ is allowed, one trivial choice is to all users keep silent with probability $\delta$, reducing the expected cost by a factor $1-\delta$. The achievable communication complexity of these methods are summarized in Table~\ref{tab:prior-work}.

Lower bound arguments typically choose an input distribution and bound the information needed for an accurate answer~\cite{braverman2016distributed}. Distance-based Fano inequalities bound the information needed to locate the aggregate within a distortion ball~\cite{duchi2013distance}. Then~\cite{fan-cover} applies this viewpoint to $f$-information for sparse aggregation under fixed-alphabet constraints. If every distortion ball has prior mass at most $p$, at least an alphabet of size $(1-\delta)/p$ is needed. Since long messages may occur rarely, a large alphabet size does not necessarily imply a large expected communication cost. 

The distance-based Fano's inequality method with aggregate counting arguments treats the users' messages jointly and does not distinguish whether they use the same or different encoders.
A complementary method relates local posterior uncertainty to squared estimation error~\cite{huang2019optimal}. They divides $d$ dimensions into blocks of equal length and chooses a prior distribution that places one active coordinate in each block. Conditional independence across users makes their local posterior variances additive, and Lemma~3.4 in~\cite{huang2019optimal} bounds this sum. This is a natural approach for mean squared error. Applying it to our accuracy requirement first converts distortion and failure probability into an average squared loss, which can weaken the resulting bound. 

\subsection{Our Contributions}
\label{sec:contributions}

We characterize the optimal communication cost for both heterogeneous and homogeneous encoder models. The proposed \emph{randomized covering} scheme allows the user $i$ choose $J_i\in\{0,\ldots,k\}$ and report $\min\{J_i,\|x_i\|_0\}$ active coordinates with their original values. The aggregate distortion is at most $D$ whenever $\sum_i J_i$ reaches $nk-\lfloor nD\rfloor$. Heterogeneous encoders can assign this total across users and use public silence with probability $\delta$. Under a common conditional rule, $J_i$'s must be independent and identically distributed after fixing the public randomness. Optimizing their encoder gives the coefficient $\Lambda_{n,k,D}(\delta)$ defined in~\eqref{eq:lambda}.

Allowing bias helps in estimating aggregation more efficiently. Unbiased sampling compensates for missing entries by increasing the reported values. When supports are disjoint, these increases add to the absolute error from omitted entries. Randomized covering keeps the reported values unchanged, so the error comes only from support omissions. It also differs from Count-Min Sketch and Bloom filters. These collision-based methods still represent every nonzero coordinate, but save bits by allowing collisions, which leads to false positive estimate. Thus, the sketches must remain detailed enough to limit such errors across a large $d$. This makes the distortion allowance less effective at reducing communication. These observations explain the smaller achievable bounds by randomized covering in Table~\ref{tab:prior-work}. Section~\ref{sec:achievability} gives geometric interpretation of randomized covering scheme and its connection with support selection.
Table~\ref{tab:prior-work} gives explicit achievable bounds under our problem setting in Section~\ref{sec:introduction}, including an explicit upper bound on the achievable communication cost under the homogeneous model. Appendix~\ref{app:quantitative-comparison} gives the derivations of our comparison.

It is worth noting the communication complexity of homogeneous model is strictly larger than that of the heterogeneous models in Table~\ref{tab:prior-work}. The homogeneous encoder model generally requires more communication because all users follow the same encoding rule and cannot coordinate their individual communication decisions. In contrast, heterogeneous encoders can distribute the required communication unevenly across users, avoiding redundant transmissions. This difference becomes particularly significant when only a small amount of information is needed to achieve the prescribed accuracy.
For example, when identifying just one active coordinate is sufficient, a heterogeneous protocol can assign this task to a single user, whereas a homogeneous protocol requires every user to independently decide whether to transmit. Section~\ref{sec:comparison} describes the cost of homogeneity under different regimes.

\begin{table}[!htbp]
\centering
\caption{Achievable expected total communication under the same problem setting, for $0\leq D<k$, $0<\delta<1$, and $d>4k$.}
\label{tab:prior-work}
\begingroup
\setlength{\tabcolsep}{4pt}
\renewcommand{\arraystretch}{1.5}
\begin{tabularx}{\linewidth}{@{}>{\raggedright\arraybackslash}p{0.26\linewidth}X@{}}
\toprule
Method & Achievable expected total communication (bits) \\
\midrule
Unbiased coordinate\newline sampling~\cite{konecny2018mean}
& $\displaystyle O\!\left((1-\delta)n\min\!\left\{k,\;k-\frac D2+\sqrt{\frac{k}{n}\ln\frac2\delta}\right\}\log d\right)$ \\
Unbiased client sampling~\cite{suresh2017mean}
& $\displaystyle O\!\left((1-\delta)n\min\!\left\{k,\;k-\frac D2+k\sqrt{\frac1n\ln\frac2\delta}\right\}\log d\right)$ \\
Count-Min~\cite{cormode2005countmin}
& $\displaystyle O\!\left((1-\delta)nk\log\frac{2nd}{\delta(1+nD)}\right)$ \\
Compressed Bloom filters~\cite{mitzenmacher2002compressed}
& $\displaystyle O\!\left((1-\delta)nk\log\frac{2nd}{\delta(1+nD)}\right)$ \\
BCH support recovery~\cite{belazzougui2024reconciliation}
& $\displaystyle O\!\left((1-\delta)nk\log d\right)$ \\
\addlinespace
Ours: heterogeneous 
& $\displaystyle (1-\delta)(nk-\lfloor nD\rfloor)\lceil\log d\rceil$ \\
Ours: homogeneous
& $\displaystyle \lambda_{n,k,D}(\delta)\leq O\!\left((1-\delta)\min\!\left\{
n(k-\lfloor D\rfloor),
nk-\lfloor nD\rfloor+\ln(2/\delta)
\right\}\log d\right)$ \\
\bottomrule
\end{tabularx}
\endgroup
\end{table}

To establish the optimality of randomized covering, we develop lower bounds based on local support identification. We prove that accurate aggregation requires each user's message to reveal sufficiently many active coordinates under an independent sparse input prior. The transmitted information are conditionally independent given public randomness. Under the homogeneous encoder constraint, it must also follow the same distribution across users, leading to a stronger communication lower bound. This argument establishes matching first-order lower bounds for both encoder models.

A related approach in~\cite{huang2019optimal} also studies the information revealed by each user's message, but differs in how this information is measured. While~\cite{huang2019optimal} uses the variance after observing the messages to bound the mean squared error, our method directly analyzes the probability that the messages identify sufficiently many active coordinates. This allows us to characterize the communication complexity under the prescribed distortion and failure probability, while capturing the additional constraint imposed by homogeneous encoding.

\begin{table}[!htbp]
\centering
\caption{Lower bounds comparison in our expected length model.  The coefficient column with $a$ means $C\geq a\log d-o(\log d)$. Here $(z)_+=\max\{z,0\}$.}
\label{tab:converse-comparison}
\begingroup
\renewcommand{\arraystretch}{1.35}
\begin{tabularx}{\linewidth}{@{}p{0.49\linewidth}p{0.28\linewidth}X@{}}
\toprule
Lower bound method & Coefficient $a$ & Encoder class \\
\midrule
Posterior variance~\cite{huang2019optimal}
& $(nk-n^2((1-\delta)D^2+2k\delta))_+$ & Both \\
Aggregate distance-based Fano~\cite{duchi2013distance,fan-cover}
& $(1-\delta)(nk-\lfloor nD\rfloor)$ & Both \\
\addlinespace
Ours: heterogeneous characterization & $(1-\delta)(nk-\lfloor nD\rfloor)$ & Heterogeneous \\
Ours: local identification under homogeneous constraint & $\Lambda_{n,k,D}(\delta)\geq (1-\delta)(nk-\lfloor nD\rfloor)$ & Homogeneous \\
\bottomrule
\end{tabularx}
\endgroup
\end{table}


Finally, we give a feasible $(1+\varepsilon)$ approximation to the optimal homogeneous policy with arithmetic complexity $O(k+f(1/\varepsilon,1/\delta)\log(n+1))$ for fixed $0<\delta<1$. A binomial policy is near optimal when the required communication complexity is large; otherwise, a finite search suffices.

The rest of this paper is organized as follows.
Section~\ref{sec:achievability} constructs the randomized covering protocol, and Section~\ref{sec:optimality} proves their first-order optimality by characterizing the lower bound results. The cost of homogeneous encoder requirement is analyzed in Section~\ref{sec:comparison}. Section~\ref{sec:computation} gives the approximation algorithm. All complete proofs can be found in the appendix.

\section{Upper Bounds via Randomized Covering}
\label{sec:achievability}

We adapt the deterministic covering principle of~\cite{fan-cover}. For a categorical input $x\in\{0,1\}^d$ with $\|x\|_0\leq k$, the basic idea of covering is to represent $x$ by a reproduction $y$ that is sufficiently close to $x$ under the prescribed distortion measure. 
From a covering perspective, each reproduction $y$ serves as a representative of a collection of input vectors lying within a prescribed distortion radius. Specifically, the distortion ball centered at $y$ contains all admissible inputs $x$ satisfying $\|x-y\|_1\le D$. This allows multiple input vectors to share the same reproduction, reducing the number of distinct representations required for communication. A smaller distortion radius provides more accurate reconstruction but generally requires a larger collection of reproductions, whereas a larger radius allows each reproduction to cover more inputs and thereby reduces the communication cost. This geometric tradeoff between covering radius and representation complexity motivates the covering scheme.

An important feature of this covering construction under histogram aggregation setting is that the distortion can be controlled through the overlap between the supports of the input and its reproduction. For binary vectors \(x_i\) and \(y_i\) with \(\supp(y_i)\subseteq\supp(x_i)\), their \(\ell_1\) distance equals the number of active coordinates omitted from the reproduction:
\[
\|x_i-y_i\|_1
=\|x_i\|_0-|\supp(x_i)\cap\supp(y_i)|.
\]
Therefore, if each user selects a covering level \(J_i\in\{0,\ldots,k\}\) and keeps exactly \(\min\{J_i,\|x_i\|_0\}\) active coordinates of its input, the resulting reproduction \(y_i\) will satisfy $\|x_i-y_i\|_1\le k-J_i$.
Thus, keeping more active coordinates corresponds to a smaller covering radius. Since the aggregate estimate is formed by averaging the individual reproductions, the triangle inequality gives $\|\hat u-u\|_1\le k-\frac1n\sum_{i=1}^n J_i$.
Consequently, the aggregate distortion is at most \(D\) whenever \(\sum_{i=1}^nJ_i\ge nk-\lfloor nD\rfloor\). This condition is also necessary when every input has exactly \(k\) active coordinates.

This covering interpretation motivates our randomized covering scheme, in which the covering radii are randomized to balance communication cost and reconstruction accuracy. 
Public randomness $R$ specifies the distribution of a covering level $J_i\in\{0,\ldots,k\}$ for every user. Conditional on $R$, each user draws its covering level using independent randomness and applies the covering operation above. The reconstruction is $\hat u=\frac1n\sum_{i=1}^n y_i$.
The message $M_i$ is the list of active indices of $y_i$, with each index represented by exactly $\lceil\log d\rceil$ bits. Consequently,
\begin{equation}
 |M_i|=\|y_i\|_0\lceil\log d\rceil\leq J_i\lceil\log d\rceil.
 \label{eq:covering-message}
\end{equation}
By the non-negativity of every $x_i-y_i$,
\begin{align}
 \|\hat u-u\|_1 =\left\|\frac1n\sum_{i=1}^n(x_i-y_i)\right\|_1 =\frac1n\sum_{i=1}^n\|x_i-y_i\|_1 \leq \frac1n\sum_{i=1}^n(k-J_i)
 =k-\frac1n\sum_{i=1}^nJ_i.
 \label{eq:covering-error}
\end{align}
Since $\sum_iJ_i$ is integer-valued, the worst-case $(D,\delta)$-accuracy condition for randomized covering is therefore equivalent to
\begin{equation}
 \Prob\!\left(\sum_{i=1}^nJ_i\geq nk-\lfloor nD\rfloor\right)\ge1-\delta.
 \label{eq:covering-feasibility}
\end{equation}
When every user is assigned level zero, in particular, we say it is a public erasure randomization since every user is instructed to choose level zero and sends an empty string, consuming no communicated bits. 

We now formulate the communication optimization for randomized covering. Let
$p_{i,R}=(p_{i,R,0},\ldots,p_{i,R,k})$ denote the conditional
distribution of $J_i$ given the public randomness $R$ at the user $i$.
The optimization takes the common form
\begin{equation}
\begin{aligned}
\min_{P_R,\,\{p_{i,R}\}_{i=1}^n}
\quad&
\E_R\left[
\sum_{i=1}^n\sum_{j=0}^k j p_{i,R,j}
\right]\\
\text{subject to}\quad&
\E_R\left[
\Pr\left(
\sum_{i=1}^n J_i
\ge nk-\lfloor nD\rfloor
\,\middle|\,R
\right)
\right]\ge 1-\delta,\\
&J_i\sim p_{i,R}
\text{ independently given }R.
\end{aligned}
\label{eq:unified-covering-optimization}
\end{equation}
The objective minimizes the expected total covering level,
while the constraint guarantees $(D,\delta)$-accuracy.



\begin{remark}[Connection to subset selection and extension to bounded inputs]
The local covering operation can in fact be interpreted as the familiar subset selection construction~\cite{konecny2018mean} for the binary input. However, the covering parlance facilitates the understanding of more general cases, e.g., $q$-ary inputs $x \in \{0,1,\ldots,q-1\}^d$ and bounded inputs $x\in[-q,q]^d$. 
As an example, for the bounded case, at a covering level $j$, one can report largest magnitude entries until the omitted absolute values sum to at most $q(k-j)$. Since omitting all but $j$ entries already meets this bound, at most $j$ entries are needed in transmission. The resulting reproduction satisfies $\|x-y\|_1\leq q(k-j)$, so aggregate accuracy follows whenever $\sum_iJ_i\geq nk-\lfloor nD/q\rfloor$.

Moreover, the covering formulation will offer a valuable insight into how we sharpen the lower bound. While randomized covering explicitly reports active coordinates to control distortion, a natural question is whether arbitrary encoding schemes must also reveal sufficient information about these coordinates. This connection motivates our local support identification approach in Section~\ref{sec:optimality}, where we establish matching communication lower bounds.
\end{remark}


Without the homogeneous constraint, heterogeneous encoding
allows $p_{i,R}$ to differ across users, so the required
total covering level can be allocated unevenly.
In this case, we will prove that the associated optimization \eqref{eq:unified-covering-optimization} has a simple answer
\begin{equation}
 \min\left\{\E\!\left[\sum_iJ_i\right]:
       \Pr\!\left(\sum_iJ_i\geq nk-\lfloor nD\rfloor\right)\geq1-\delta\right\}
       =(1-\delta)(nk-\lfloor nD\rfloor).
 \label{eq:heterogeneous-level-optimization}
\end{equation}

\begin{lemma}[Heterogeneous upper bound]
\label{lem:het-achievability}
For every $d\ge\max\{2,k\}$,
\begin{equation*}
 \chet(D,\delta)\leq (1-\delta)(nk-\lfloor nD\rfloor)\lceil\log d\rceil.
\end{equation*}
\end{lemma}

When the protocol does not choose erasure, the users can send exactly $nk-\lfloor nD\rfloor$ active indices in the worst case. Inputs with fewer active coordinates may require fewer bits. For example, when $nk-\lfloor nD\rfloor=1$, one user sends at most one active index while all others remain silent. The lower bound in Section~\ref{sec:het-converse} shows that this expected communication cost is optimal to first order among all heterogeneous protocols.


Different from the heterogeneous case, in \eqref{eq:unified-covering-optimization}, homogeneous encoding imposes $p_{1,R}=\cdots=p_{n,R}=p_R$ for every realization of $R$. Consequently, the covering levels are conditionally i.i.d., and their sum may fluctuate around the required threshold. Thus, the homogeneous model has a more restricted feasible set. 

Write $p_R=(p_{R,0},\ldots,p_{R,k})$. The homogeneous specialization of \eqref{eq:unified-covering-optimization} is
\begin{equation}
 \begin{aligned}
 \Lambda_{n,k,D}(\delta):={}&
 \min_{\mathcal L(R),\,p_R}\ n\E_R\!\left[\sum_{j=0}^k jp_{R,j}\right]\\
 \text{subject to}\quad&
 \E_R\!\left[\Pr_{J_1,\ldots,J_n\,\mathrm{i.i.d.}\sim p_R}
 \!\left\{\sum_{i=1}^nJ_i\geq nk-\lfloor nD\rfloor\right\}\right]\geq1-\delta.
 \end{aligned}
 \label{eq:lambda}
\end{equation}
Apply the covering construction with a minimizing policy. The message-length bound \eqref{eq:covering-message} gives
\begin{align*}
 \sup_{\|x_i\|_0\leq k}\E\!\left[\sum_{i=1}^n|M_i|\right]\le\sum_{i=1}^n\E[J_i]\lceil\log d\rceil=n\E_R\!\left[\sum_{j=0}^k jp_{R,j}\right]\lceil\log d\rceil=\Lambda_{n,k,D}(\delta)\lceil\log d\rceil.
\end{align*}
This particular protocol gives the upper bound
\begin{equation}
 \chom(D,\delta)\le
 \Lambda_{n,k,D}(\delta)\lceil\log d\rceil.
 \label{eq:rce_hom_upper}
\end{equation}
To understand the optimization, compare two simple strategies. A deterministic common level must round the quota up separately at every user. Independent random levels can reduce that rounding cost when their sum concentrates near its mean. The following bounds quantify both possibilities.

\begin{theorem}[Bounds on the homogeneous coefficient]
\label{thm:bounds}
For $0\leq D<k$ and $0\le\delta<1$,
\begin{equation}
 (1-\delta)(nk-\lfloor nD\rfloor)
 \le\Lambda_{n,k,D}(\delta)
 \le(1-\delta)n(k-\lfloor D\rfloor).
 \label{eq:elementary_bounds}
\end{equation}
If $0<\delta<1$ and $nk-\lfloor nD\rfloor\geq \max\{2,1/\delta\}$, then
\begin{equation}
\Lambda_{n,k,D}(\delta) \leq (1-\delta)(nk-\lfloor nD\rfloor)\left[1+O\left(\sqrt{\frac{\ln(nk-\lfloor nD\rfloor)}{nk-\lfloor nD\rfloor}}\right)\right].
\label{eq:concentration_bound}
\end{equation}
\end{theorem}

Combining the coefficient bounds with \eqref{eq:rce_hom_upper} gives $\chom \leq (1-\delta)n(k-\lfloor D\rfloor)\lceil\log d\rceil$
and, for $0<\delta<1$ and $nk-\lfloor nD\rfloor\geq\max\{2,1/\delta\}$,
\begin{equation*}
    \chom \leq (1-\delta)(nk-\lfloor nD\rfloor)\left[1+O\left(\sqrt{\frac{\ln(nk-\lfloor nD\rfloor)}{nk-\lfloor nD\rfloor}}\right)\right]\times \lceil \log d\rceil.
\end{equation*}

\section{Lower Bounds on Communication}
\label{sec:optimality}

The covering constructions allocate local coordinate counts, whereas the model permits arbitrary messages and estimators. We first show that accurate aggregation requires identifying enough active coordinates. This common bound gives the heterogeneous characterization by a total-count argument and the homogeneous characterization under the common encoder restriction.


Let $X_1,\ldots,X_n$ be independent and uniform over binary vectors with exactly $k$ active coordinates. Expected communication under this prior lower bounds worst-case cost. For $0<\tau<1$, define the posterior lists and their true-support counts by
\begin{equation}
 I_i=\{a:\Prob(X_i(a)=1\mid M_i,R)>\tau\},\qquad \text{and} \qquad
 J_i=|\supp(X_i)\cap I_i|.
 \label{eq:identified-counts}
\end{equation}
These are analysis variables; neither the encoder nor the decoder must compute them. The list contains coordinates with sufficiently large local posterior probability. 

\begin{lemma}[Accuracy requires local support identification]
\label{lem:identification}
For any $(D,\delta)$-accurate protocol in either encoder model, the counts in \eqref{eq:identified-counts} are independent conditional on $R$ and satisfy
\begin{equation}
 \Prob\!\left(\sum_{i=1}^nJ_i<nk-\lfloor nD\rfloor\right)
 \leq\delta+\frac{n^2\tau(k+D)}{\lfloor nD\rfloor+1-nD}.
 \label{eq:identification-success}
\end{equation}
With $\tau=1/\log d$ and fixed $n,k$, for all sufficiently large $d$,
\begin{equation}
 \E\!\left[\sum_{i=1}^n|M_i|\right]
 \geq \E\!\left[\sum_{i=1}^nJ_i\right]\log d-O(\log\log d).
 \label{eq:identified-length}
\end{equation}
\end{lemma}
\begin{proof}[Proof sketch]
Conditional on $R$, the pairs $(X_i,M_i)$ are independent, so $J_i$ are also independent. If $\sum_iJ_i<nk-\lfloor nD\rfloor$ but the estimate is accurate, the active coordinates outside the posterior lists must satisfy
\[
 \sum_{i=1}^n\sum_{a=1}^d\hat u(a)
 \mathds{1}\{X_i(a)=1,\ a\notin I_i\}
 \geq\frac{\lfloor nD\rfloor+1-nD}{n}.
\]
Each coordinate outside $I_i$ has posterior probability at most $\tau$. Thus, the expected left-hand side is at most $n\tau(k+D)$. Markov's inequality, together with the allowed failure probability $\delta$, gives \eqref{eq:identification-success}.

For the communication bound, $|I_i|\leq k/\tau$. Under the uniform sparse prior, a fixed list of this size contains at least $j$ true active coordinates with probability at most $(k/(\tau\lfloor d/k\rfloor))^j$. There are fewer than $2^{jL+1}$ binary strings of length below $jL$, so
\[
 \Prob(J_i\geq j,\ |M_i|<jL\mid R)
 \leq2\left(\frac{2^Lk}{\tau\lfloor d/k\rfloor}\right)^j.
\]
Summing over users and $j$ gives \eqref{eq:length-count-general}. Taking $\tau=1/\log d$ and $L=\log[d/(8nk^2(\log d)^2)]$ makes the total exceptional probability $O(1/\log d)$ and gives \eqref{eq:identified-length}. Appendix~\ref{app:proof-identification} gives the complete proof.
\end{proof}

\label{sec:het-converse}

Heterogeneous encoders do not require the conditional distributions of $J_i$ given $R$ to be identical. The success probability in Lemma~\ref{lem:identification} therefore implies that any heterogeneous encoder satisfying the accuracy requirement must identify, on average, at least as many active coordinates as required by the optimal heterogeneous covering scheme.

\begin{theorem}[Heterogeneous communication characterization]
\label{thm:heterogeneous}
For fixed $n,k,D,\delta$ with $0\leq D<k$ and $0\leq\delta<1$,
\begin{equation}
 \chet(D,\delta)=(1-\delta)(nk-\lfloor nD\rfloor)\log d
 +O(\log\log d),\qquad d\to\infty.
 \label{eq:het-range-characterization}
\end{equation}
The following explicit bound also holds for every $d\geq\max\{2,nk\}$:
\begin{align}
 \chet(D,\delta)\geq{}&(1-\delta)(nk-\lfloor nD\rfloor)\log d
 -nk\log(2nk)-1 -n\log\left(e(k\lceil\log d\rceil+1)\right).
 \label{eq:finite-het-converse}
\end{align}
\end{theorem}

Thus assigning just enough total level $\sum_i J_i$ with probability $1-\delta$ is optimal to first order among arbitrary heterogeneous encoders. 
For a homogeneous protocol, the distribution of $J_i$ supplied by Lemma~\ref{lem:identification} are conditionally i.i.d., and this additional constraint strengthens the lower bound.

\begin{theorem}[Homogeneous communication characterization]
\label{thm:main}
Fix $n,k,D,\delta$ with $0\leq D<k$ and $0\leq\delta<1$. For all sufficiently large $d$,
\begin{align}
 \chom(D,\delta)\geq{}&
 \Lambda_{n,k,D}\!\left(\delta+
 \frac{n^2(k+D)}{(\lfloor nD\rfloor+1-nD)\log d}\right)\log d
 -O(\log\log d).
 \label{eq:hom-finite-characterization}
\end{align}
Together with \eqref{eq:rce_hom_upper}, this gives
\begin{equation}
 \chom(D,\delta)=\Lambda_{n,k,D}(\delta)\log d+o(\log d),
 \qquad d\to\infty.
 \label{eq:hom-range-characterization}
\end{equation}
\end{theorem}

The two characterizations come from the same local bound but with different distribution constraint on $J_i$'s. Compactness removes the posterior-threshold relaxation in the leading coefficient. We next compare the resulting coefficients.

\section{The Communication Cost of Homogeneity}
\label{sec:comparison}

The preceding sections reduce the comparison of arbitrary protocols to a comparison of two leading coefficients. In this section, we compare the communication costs of homogeneous and heterogeneous encoders to determine when using the common encoding rule increases communication and by how much.

\begin{theorem}[When homogeneity has no leading cost]
\label{thm:comparison}
For every $0\leq D<k$ and $0\le\delta<1$,
\begin{equation*}
 1\le\frac{\Lambda_{n,k,D}(\delta)}
 {(1-\delta)(nk-\lfloor nD\rfloor)}
 \le\frac{n(k-\lfloor D\rfloor)}{nk-\lfloor nD\rfloor}
 \leq n.
\end{equation*}
The equality $\Lambda_{n,k,D}(\delta) = (1-\delta)(nk-\lfloor nD\rfloor)$ holds if and only if $\lfloor nD\rfloor=n\lfloor D\rfloor$.
\end{theorem}


When the required communication cost grows, concentration makes the common stochastic encoder nearly as efficient as a fixed heterogeneous encoder. When $D$ is close to $k$, heterogeneous encoders can meet the accuracy requirement with only a few users transmitting. Requiring all users to use the same encoding rule can then substantially increase communication. The next theorem makes these two behaviors precise. 

\Needspace{7\baselineskip}
\begin{theorem}[Small and large costs of homogeneity]
\label{thm:homogeneity-regimes}
For fixed $n,k,D,\delta$, the communication characterizations in Section~\ref{sec:optimality} give, as $d\to\infty$,
\begin{equation}
 \frac{\chom(D,\delta)}{\chet(D,\delta)}
  = \frac{\Lambda_{n,k,D}(\delta)}{(1-\delta)(nk-\lfloor nD\rfloor)}+o(1).
 \label{eq:homogeneity-fixed-parameter-ratio}
\end{equation}
Suppose
\[
 nk-\lfloor nD\rfloor\to \infty,
 \qquad
 \frac{nk\log(2nk)+1+n\log\left(e(k\lceil\log d\rceil+1)\right)}
      {(nk-\lfloor nD\rfloor)\log d}\to 0.
\]
 Then,
\begin{equation}
 \chom(D,\delta)\sim\chet(D,\delta)
 \sim(1-\delta)(nk-\lfloor nD\rfloor)\log d.
 \label{eq:ratio_bound_one}
\end{equation}
Suppose distortion $D$ satisfies 
\begin{equation}
 k-\frac1n\leq D<k.
 \label{eq:ratio-n-distortion}
\end{equation}
For fixed $n,k,D,\delta$, as $d\to\infty$,
\begin{equation}
 \frac{\chom(D,\delta)}{\chet(D,\delta)}=\frac{n(1-\delta^{1/n})}{1-\delta}+o(1),
 \label{eq:ratio-n}
\end{equation}
which approaches $n$ as $\delta\rightarrow 0$.
\end{theorem}

\begin{proof}[Proof sketch]
For fixed parameters, dividing the homogeneous and heterogeneous communication characterizations gives \eqref{eq:homogeneity-fixed-parameter-ratio}. For fixed $0<\delta<1$, the concentration bound in Theorem~\ref{thm:bounds} implies, as $nk-\lfloor nD\rfloor\to\infty$,
\[
 \Lambda_{n,k,D}(\delta)
 =(1-\delta)(nk-\lfloor nD\rfloor)(1+o(1)).
\]
Combine the homogeneous covering upper bound with $\chet\leq\chom$ and the finite heterogeneous lower bound \eqref{eq:finite-het-converse}. The stated remainder condition makes the error in that lower bound negligible relative to $(1-\delta)(nk-\lfloor nD\rfloor)\log d$, proving \eqref{eq:ratio_bound_one}.

When $k-1/n\leq D<k$, the required total level is one. Replacing each positive $J_i$ by one preserves success and cannot increase the cost, so it suffices to consider Bernoulli levels with common conditional probability $p_{R,1}$. Feasibility and Jensen's inequality give
\[
 (1-\E_R[p_{R,1}])^n
 \leq\E_R[(1-p_{R,1})^n]\leq\delta.
\]
Consequently, the expected total level is at least $n(1-\delta^{1/n})$. The constant choice $p_{R,1}=1-\delta^{1/n}$ attains this bound, proving $\Lambda_{n,k,D}(\delta)=n(1-\delta^{1/n})$. Substitution into \eqref{eq:homogeneity-fixed-parameter-ratio} gives \eqref{eq:ratio-n}, and letting $\delta\to 0$ gives the limit $n$. Appendix~\ref{app:proof-homogeneity-regimes} gives the complete proof.
\end{proof}



\section{Efficient Computation of the Optimal Covering}
\label{sec:computation}

The characterization in \eqref{eq:lambda} leaves an algorithmic question: how can we find a nearly optimal common policy? A public choice $p_R$ is itself a distribution on the $k+1$ levels. We first reduce this infinite-dimensional search to a mixture of two distributions, then discretize the individual distributions without sacrificing feasibility.

\begin{lemma}[Two public randomizations suffice]
\label{lem:two-public-modes}
For every $0\leq D<k$ and $0\le\delta<1$, the minimum in \eqref{eq:lambda} can be attained by a mixture of at most two conditional distributions $p_r,p_{r'}$. Within each mode, the common distribution of the $J_i$ may be supported on
\[
 \{0,\ldots,\min\{k,nk-\lfloor nD\rfloor\}\}.
\]
When $0<\delta<1$, the mixture can be chosen to have success probability exactly $1-\delta$. 
\end{lemma}

Although two-point public randomization suffices, each conditional distribution may involve multiple covering levels, making the computation of $\Lambda_{n,k,D}(\delta)$ nontrivial. To obtain an approximation $\hat{\Lambda}_{n,k,D(\delta)}$ in the form of \eqref{eq:eptas}, we develop an algorithm that combines a binomial construction for small distortion with an enumeration of distributions over restricted supports in the remaining cases. This approach avoids a direct discretization of all $(k+1)$ probabilities and provides an approximation guarantee with an explicit computational complexity.

\begin{center}
\begin{minipage}{0.97\linewidth}
\small\hrule\vspace{0.5em}
\noindent\textbf{Algorithm 1: Efficient homogeneous covering approximation}

\smallskip
\noindent\textbf{Input:} $n,k,D$, rational $0<\delta<1$, and $0<\varepsilon<1$.

\noindent\textbf{Output:} A feasible mixture of $p_r$ and $p_{r'}$ and its expected total covering level $\widehat\Lambda_{n,k,D}(\delta)$.

\smallskip
\noindent\textbf{Set} $t=nk-\lfloor nD\rfloor$, $h=\min\{k,t\}$,
$L=\lceil\log(8/(\varepsilon\delta))\rceil$, and
$B=(4t+8\lceil\log(2/\delta)\rceil)/\delta$.
\begin{enumerate}
\item If $t\ge64L/\varepsilon^2$, set
\[
 p=\min\left\{1,\frac{t+2L+2\lceil\sqrt{tL}\rceil}{nk}\right\}.
\]
Return $J_i\sim\operatorname{Bin}(k,p)$ with public probability
$(1-\delta)/(1-\varepsilon\delta/8)$, and the all-zero erasure randomization with the remaining probability. Stop.

\item Otherwise, enumerate probability vectors on $\{0,\ldots,h\}$ as follows:
\begin{itemize}
\item If $n\le B+\varepsilon(1-\delta)t/2$, set
$m=\lceil nh/[\varepsilon(1-\delta)t]\rceil$ and enumerate all
$p$ with $mp_j\in\mathbb Z_{\ge0}$ and $\sum_jp_j=1$.
\item If $n>B+\varepsilon(1-\delta)t/2$, set
$\rho=\varepsilon(1-\delta)t/[h(h+1)]$ and enumerate
\[
 p_j=\frac{a_j\rho}{n}\ (1\le j\le h),\qquad
 p_0=1-\sum_{j=1}^h p_j\ge0,
\]
with integers $0\le a_j\le1+\lceil B/\rho\rceil$.
\end{itemize}

\item For each candidate $p$, compute $\Pr_p(\sum_iJ_i\ge t)$
and $n\sum_jjp_j$. Return the minimum cost feasible point on
the lower convex hull of these success--cost pairs, as a mixture of at most two distributions.
\end{enumerate}
\vspace{0.1em}\hrule
\end{minipage}
\end{center}

The guarantee below preserves the original accuracy requirement. Its running time bound concerns arithmetic operations. 

\begin{theorem}[Efficient approximation of the optimal policy]
\label{thm:eptas}
Fix $0<\delta<1$. For every $0<\varepsilon<1$, Algorithm~1 returns a feasible homogeneous randomized covering public mixture with $\widehat\Lambda_{n,k,D}(\delta)$ satisfying
\begin{equation}
 \Lambda_{n,k,D}(\delta)\le\widehat\Lambda_{n,k,D}(\delta)
 \le(1+\varepsilon)\Lambda_{n,k,D}(\delta)
 \label{eq:eptas}
\end{equation}
with arithmetic complexity
\[
O\left(k+f(1/\varepsilon,1/\delta)\log(n+1)\right),
\]
where
\[
f\left(\frac{1}{\varepsilon},\frac{1}{\delta}\right)
=
\exp\left[O\left(
\frac{\log(8/(\varepsilon\delta))}{\varepsilon^2}
\log\left(
\frac{\log^2(8/(\varepsilon\delta))}
{\delta(1-\delta)\varepsilon^5}
\right)\right)\right].
\]

\end{theorem}


\section{Conclusion}
We study the communication complexity of distributed sparse histogram estimation under prescribed $\ell_1$ distortion and failure probability constraints. Our randomized covering scheme chooses how accurately each user represents its input and optimizes these choices to reduce communication cost. 
For heterogeneous encoders, the scheme achieves an expected total communication cost of at most $(1-\delta)(nk-\lfloor nD\rfloor)\lceil\log d\rceil$ bits, while for homogeneous encoders, the corresponding upper bound is $\Lambda_{n,k,D}(\delta)\lceil\log d\rceil$ bits, where $(1-\delta)(nk-\lfloor nD\rfloor)
\leq \Lambda_{n,k,D}(\delta)
\leq (1-\delta)n(k-\lfloor D\rfloor)$.
Our achievability results improve upon those of existing methods. Specifically, the communication complexity of unbiased coordinate and client sampling are no better than $O((1-\delta)n\log d\min\{k, k-\frac{D}{2}+\sqrt{\frac{1}{n}\ln\frac2\delta}\})$, and Count-Min Sketch and compressed Bloom filters achieve $O((1-\delta)nk\log\frac{2nd}{\delta(1+nD)})$ bits.

The matching lower bound is established by a local support identification, yielding a first-order characterization of communication complexity together with the lower bound.
The resulting characterization also quantifies how the communication advantage of heterogeneous over homogeneous encoders varies across parameter regimes. The observation clarifies how sparsity, distortion, failure probability, and encoder heterogeneity jointly determine the communication required for distributed sparse histogram estimation.

\section*{AI Statement}
The authors used ChatGPT (OpenAI) to improve the clarity and presentation of the manuscript and to assist in improving the computational efficiency of Algorithm 1. The authors independently verified the resulting algorithm and its theoretical guarantees and take full responsibility for the correctness and originality of the work. The tool materially affected Section~\ref{sec:computation}. More details can be found in Appendix~\ref{app:computation-proofs}. The authors verified the correctness and originality of all content.

\newpage

\appendices

\section{Omitted Proofs in Section~\ref{sec:achievability}}
\label{app:achievability-proofs}

\subsection{Proof of Lemma~\ref{lem:het-achievability}}
\label{app:proof-het-achievability}
\begin{proof}
Choose integers $j_1,\ldots,j_n\in\{0,\ldots,k\}$ satisfying
\[
 \sum_{i=1}^n j_i=nk-\lfloor nD\rfloor.
\]
Such an allocation exists because $0<nk-\lfloor nD\rfloor\leq nk$. With public probability $1-\delta$, user $i$ employs covering level $j_i$. With the remaining probability $\delta$, all users send empty strings. In the first case, \eqref{eq:covering-error} implies
\begin{align*}
 \|\hat u-u\|_1 \leq k-\frac1n\sum_{i=1}^n j_i = \frac{\lfloor nD\rfloor}{n}\leq D.
\end{align*}
The protocol is therefore $(D,\delta)$-accurate. Each message has at most $j_i$ indices in the active randomization and zero in the erasure mode, giving
\begin{align*}
 \sup_{\|x_i\|_0\leq k}\E\!\left[\sum_i|M_i|\right]
 \le(1-\delta)\sum_{i=1}^n j_i\lceil\log d\rceil
 =(1-\delta)(nk-\lfloor nD\rfloor)\lceil\log d\rceil.
\end{align*}
The allocation $j_i$ may differ between users conditional on the same public randomization, so the construction is heterogeneous unless all $j_i$ are equal.
Conversely, any feasible allocation satisfies
\[
 \E\sum_iJ_i\geq(nk-\lfloor nD\rfloor)
 \Pr\!\left(\sum_iJ_i\geq nk-\lfloor nD\rfloor\right)
 \geq(1-\delta)(nk-\lfloor nD\rfloor).
\]
The construction attains this value and proves \eqref{eq:heterogeneous-level-optimization}.
\end{proof}

\subsection{Proof of Theorem~\ref{thm:bounds}}
\label{app:proof-bounds}
\begin{proof}
The lower bound follows because the total level is at least the threshold on every successful execution:
\[
 \E\sum_iJ_i\geq(nk-\lfloor nD\rfloor)
 \Prob\!\left(\sum_iJ_i\geq nk-\lfloor nD\rfloor\right)
 \geq(1-\delta)(nk-\lfloor nD\rfloor).
\]
For the first upper bound, all users choose level $k-\lfloor D\rfloor$ with public probability $1-\delta$, and all choose zero otherwise. The active randomization succeeds deterministically and has total level $n(k-\lfloor D\rfloor)$.

For the sharper bound, let the active randomization use independent levels $J_i\sim\operatorname{Bin}(k,p)$. Their sum has distribution $\operatorname{Bin}(nk,p)$ and mean $nkp$. We place this mean slightly above the threshold, so concentration makes the active randomization reliable. For any $v>0$, the exponential-moment argument for the Chernoff bound~\cite[Chapter~2]{boucheron2013concentration} gives
\begin{align*}
 \Prob\!\left(\sum_iJ_i<nk-\lfloor nD\rfloor\right)
 &\leq e^{v(nk-\lfloor nD\rfloor)}(1-p+pe^{-v})^{nk}\\
 &\leq\exp\!\left(-v[nkp-(nk-\lfloor nD\rfloor)]+
                         \frac{nkp\,v^2}{2}\right).
\end{align*}
Here we used $1+x\leq e^x$ and $e^{-v}\leq1-v+v^2/2$.
When $nkp>nk-\lfloor nD\rfloor$, minimizing the displayed quadratic gives the upper bound
\[
 \exp\!\left(-\frac{[nkp-(nk-\lfloor nD\rfloor)]^2}{2nkp}\right).
\]
Consequently, an active-randomization failure probability at most $0<\beta\leq\delta$ is ensured by
\begin{equation}
 nkp\geq nk-\lfloor nD\rfloor+\ln\frac1\beta+
 \sqrt{\left(\ln\frac1\beta\right)^2+
             2(nk-\lfloor nD\rfloor)\ln\frac1\beta}.
 \label{eq:binomial_sufficient_condition}
\end{equation}
Choose $p$ to make this an equality if possible, and set $p=1$ otherwise. In the latter case the active randomization succeeds with probability one. In either case,
\[
 nkp\leq nk-\lfloor nD\rfloor+2\ln\frac1\beta+
             2\sqrt{(nk-\lfloor nD\rfloor)\ln\frac1\beta}.
\]

Use the active randomization with public probability $(1-\delta)/(1-\beta)$ and erase otherwise. This probability is at most one, and the overall success probability is at least $1-\delta$. Its expected total level gives
\[
 \Lambda_{n,k,D}(\delta)\leq\frac{1-\delta}{1-\beta}
 \left[nk-\lfloor nD\rfloor+2\ln\frac1\beta+
       2\sqrt{(nk-\lfloor nD\rfloor)\ln\frac1\beta}\right].
\]
Finally, take $\beta=1/(nk-\lfloor nD\rfloor)$, which is admissible under the stated hypothesis. Dividing by $(1-\delta)(nk-\lfloor nD\rfloor)$ gives
\[
 \frac{\Lambda_{n,k,D}(\delta)}{(1-\delta)(nk-\lfloor nD\rfloor)}
 \leq
 \frac{1+2\frac{\ln(nk-\lfloor nD\rfloor)}{nk-\lfloor nD\rfloor}
       +2\sqrt{\frac{\ln(nk-\lfloor nD\rfloor)}{nk-\lfloor nD\rfloor}}}
      {1-\frac1{nk-\lfloor nD\rfloor}}.
\]
The right-hand side is $1+O(\sqrt{\ln(nk-\lfloor nD\rfloor)/(nk-\lfloor nD\rfloor)})$, with a universal implied constant. This proves \eqref{eq:concentration_bound}.
\end{proof}

\section{Omitted Proofs in Section~\ref{sec:optimality}}
\label{app:optimality-proofs}

\subsection{Proof of Lemma~\ref{lem:identification}}
\label{app:proof-identification}
\begin{proof}
We may first modify the decoder without changing its communication cost or
reducing its accuracy. Clip each output coordinate to $[0,1]$, which cannot
increase its distance to any admissible $u$. If the clipped output has
$\ell_1$ norm greater than $k+D$, replace it by zero. Such an output cannot
have been successful, since $\|u\|_1\leq k$ and success implies
$\|\hat u\|_1\leq k+D$. Hence we may assume, on every outcome, that
\[
 \hat u\geq0,\qquad \|\hat u\|_1\leq k+D.
\]
Suppose $j = \sum_{i=1}^n \mathds{1}\{X_i(a)=1, a\notin I_i\}$, i.e., a coordinate $a$ is truly active for exactly \(j\) users whose posterior candidate sets do not contain \(a\). When $j\geq1$, we have $u(a)\geq j/n$ and
\begin{align*}
 |u(a)-\hat u(a)|&\geq u(a)-\hat u(a)
 \ge\frac jn-\hat u(a)\ge\frac jn-j\hat u(a)\\
 & = \frac{1}{n}\sum_{i=1}^n \mathds{1}\{X_i(a)=1, a\notin I_i\}  - \sum_{i=1}^n \hat{u}(a) \mathds{1}\{X_i(a)=1, a\notin I_i\}.
\end{align*}
When $j=0$, the last expression is zero, so the same final lower bound follows from $|u(a)-\hat u(a)|\geq0$. Since each $X_i$ has exactly $k$ active coordinates, $k - J_i = \sum_{a=1}^d \mathds{1}\{X_i(a)=1, a\notin I_i\}$. 
Summing the coordinatewise lower bounds gives
\begin{align*}
 \|u-\hat u\|_1 =\frac{nk-\sum_iJ_i}{n}-\sum_{i=1}^n\sum_{a=1}^d
       \hat u(a)\mathds{1}\{X_i(a)=1,\ a\notin I_i\}.
\end{align*}
On the event $\sum_iJ_i<nk-\lfloor nD\rfloor$ and $\|u-\hat u\|_1\leq D$, integrality implies that
\begin{align*}
 \sum_{i=1}^n\sum_{a=1}^d
       \hat u(a)\mathds{1}\{X_i(a)=1,\ a\notin I_i\} &\ge\frac{nk-\sum_iJ_i}{n}-\|u-\hat u\|_1\\
 &\ge\frac{\lfloor nD\rfloor+1}{n}-D
       =\frac{\lfloor nD\rfloor+1-nD}{n}.
\end{align*}
Conditional independence under the product prior gives
\[
 \Prob(X_i(a)=1\mid M_1,\ldots,M_n,R)
 =\Prob(X_i(a)=1\mid M_i,R).
\]
Conditioning first on all messages and $R$ makes both $\hat u$ and the sets
$I_i$ fixed. Therefore,
\begin{align*}
 \E\!\left[\sum_{i=1}^n\sum_{a=1}^d
       \hat u(a)\mathds{1}\{X_i(a)=1,\ a\notin I_i\}\right]
       &=\E\!\left[\sum_{i=1}^n\sum_{a\notin I_i}
       \hat u(a)\E\left[\mathds{1}\{X_i(a)=1\}\mid M_1,\ldots,M_n,R\right]\right]\\
       &=\E\!\left[\sum_{i=1}^n\sum_{a\notin I_i}
       \hat u(a)\Prob(X_i(a)=1\mid M_i,R)\right]\\
 &\le\tau\E\!\left[\sum_{i=1}^n\sum_{a\notin I_i}\hat u(a)\right]
 \leq n\tau\E\|\hat u\|_1\leq n\tau(k+D).
\end{align*}
Combining the event inclusion above with Markov's inequality gives
\begin{align*}
 \Prob\!\left(\sum_iJ_i<nk-\lfloor nD\rfloor\right) \le\delta+\frac{n^2\tau(k+D)}{\lfloor nD\rfloor+1-nD}.
\end{align*}

To transmit the identified coordinates, we use the local uniform prior and count short binary strings. Fix a public realization $R=r$ and a binary string $m$; lists for messages of conditional probability zero may be taken empty. The posterior identification set satisfies
\begin{align*}
 \tau|I_i(m,r)| \le\sum_{a\in I_i(m,r)}\Prob(X_i(a)=1\mid M_i=m,R=r) \le\sum_{a=1}^d\Prob(X_i(a)=1\mid M_i=m,R=r)=k,
\end{align*}
implying that $|I_i(m,r)|\leq k/\tau$. For any fixed $j$ coordinates, the uniform prior gives the probability that all of these $j$ coordinates are active, which is
\[
 \frac{\binom{d-j}{k-j}}{\binom dk}
 =\prod_{\ell=0}^{j-1}\frac{k-\ell}{d-\ell}
 \le\left(\frac kd\right)^j \leq \frac{1}{\lfloor d/k\rfloor^j}.
\]
A union bound over the subsets of $I_i(m,r)$ therefore shows that, for $1\leq j\leq k$,
\begin{align*}
 \Prob\!\left(|\supp(X_i)\cap I_i(m,r)|\geq j\right) \leq\binom{|I_i(m,r)|}{j}\frac{\binom{d-j}{k-j}}{\binom dk}
 \leq \frac{|I_i(m,r)|^j}{j!}\cdot\frac{1}{\lfloor d/k\rfloor^j}
 \leq \frac{(k/\tau)^j}{j!}\cdot\frac{1}{\lfloor d/k\rfloor^j}
 \leq\left(\frac{k}{\tau\lfloor d/k\rfloor}\right)^j.
\end{align*}

For a positive length $L$, the number of binary strings shorter than $jL$ is $\sum_{\ell=0}^{\lceil jL\rceil-1}2^\ell<2^{jL+1}$.
Summing over these strings gives
\begin{align*}
 \Prob(J_i\geq j,\ |M_i|<jL\mid R=r) \le2\left(\frac{2^Lk}{\tau\lfloor d/k\rfloor}\right)^j.
\end{align*}
Since $J_i$ is integer-valued,
\begin{align*}
 J_i =\sum_{j=1}^k \mathds{1}\{J_i\geq j,\ jL\le|M_i|\}
 +\sum_{j=1}^k \mathds{1}\{J_i\geq j,\ |M_i|<jL\}
 \le\frac{|M_i|}{L}
 +\sum_{j=1}^k \mathds{1}\{J_i\geq j,\ |M_i|<jL\}.
\end{align*}
Taking expectations and applying the preceding inequality gives
\begin{equation}
 \E\!\left[\sum_{i=1}^n|M_i|\right]
 \geq L\left(\sum_{i=1}^n\E[J_i]
 -2n\sum_{j=1}^k
 \left(\frac{2^Lk}{\tau\lfloor d/k\rfloor}\right)^j\right).
 \label{eq:length-count-general}
\end{equation}

Choose  \[
 \tau = \frac{1}{\log d}\qquad \text{and}\qquad L=\log\frac{d}{8nk^2(\log d)^2}.
\]
For fixed $n,k$ and sufficiently large $d$, we have $L>0$ and $\lfloor d/k\rfloor\geq d/(2k)$. Hence, $(2^Lk)/(\tau\lfloor d/k\rfloor)\leq 1/(4n\log d)$ and consequently $2n\sum_{j=1}^k
 \left(\frac{2^Lk}{\tau\lfloor d/k\rfloor}\right)^j \leq \frac1{\log d}$.
Substituting the above inequality into \eqref{eq:length-count-general} implies that, for either encoder model,
\begin{equation}
 \E\!\left[\sum_{i=1}^n|M_i|\right]
 \ge\left(\sum_{i=1}^n\E[J_i]\right)\log d
 -O(\log\log d).
 \label{eq:hom-length-converse}
\end{equation}
The hidden constant depends only on $n,k$, since $\sum_i\E[J_i]\leq nk$. This proves \eqref{eq:identified-length} and completes the common bound.
\end{proof}

\subsection{Proof of Theorem~\ref{thm:heterogeneous}}
\label{app:proof-heterogeneous}
\begin{proof}
Apply Lemma~\ref{lem:identification} with $\tau=1/\log d$. Multiplying the success probability in \eqref{eq:identification-success} by the required total level gives
\[
 \E\sum_iJ_i\geq(1-\delta)(nk-\lfloor nD\rfloor)
 -\frac{n^2(k+D)(nk-\lfloor nD\rfloor)}
 {(\lfloor nD\rfloor+1-nD)\log d}.
\]
For fixed $n,k,D,\delta$, substitution into \eqref{eq:identified-length} gives
\[
 \chet(D,\delta)\geq(1-\delta)(nk-\lfloor nD\rfloor)\log d
 -O(\log\log d).
\]
The worst-case cost is at least the prior-averaged cost for each protocol, so taking the infimum preserves this bound. Lemma~\ref{lem:het-achievability} supplies the opposite bound, with an $O(1)$ rounding correction. This proves \eqref{eq:het-range-characterization}.

We prove the explicit bound \eqref{eq:finite-het-converse} separately to keep a remainder valid when the parameters grow.

We first bound the number of aggregates that a successful estimate can leave unresolved, then translate the information revealed into expected message length. Choose $U$ uniformly among the $nk$-subsets of $\{1,\ldots,d\}$. Partition each chosen set into $n$ disjoint supports of size $k$ and give their indicator vectors to the users. Their aggregate is $\mathds{1}_U/n$. Thus,
\[
 H(U\mid R)=H(U)=\log\binom d{nk}.
\]
The worst-case accuracy requirement gives success probability at least $1-\delta$ under this prior.

Fix a decoder output $\hat u$. If at least one $nk$-subset has aggregate within distance $D$ of $\hat u$, choose such a set $U$ as a reference. For any other such subset $V$ with exactly $nk$ elements whose corresponding aggregate is also within distance $D$ of the same reconstruction $\hat u$,
\begin{align*}
 \frac{2|U\setminus V|}{n} = \left\|\frac{\mathds{1}_U}{n}-\frac{\mathds{1}_V}{n}\right\|_1 \leq \left\|\frac{\mathds{1}_U}{n}-\hat u\right\|_1 + \left\|\hat u-\frac{\mathds{1}_V}{n}\right\|_1 \leq 2D.
\end{align*}
Since $|U\setminus V|$ is an integer, it is at most $\lfloor nD\rfloor$. At a distance $|U\setminus V| = j\leq \lfloor nD\rfloor$, choose the $j$ deleted elements from $U$ and the $j$ added elements from its complement. Therefore,
\begin{align}
 \left|\left\{V:|V|=nk,\ \left\|\frac{\mathds{1}_V}{n}-\hat u\right\|_1\leq D\right\}\right| \leq \sum_{j=0}^{\lfloor nD\rfloor}\binom{nk}{j}\binom{d-nk}{j}
 \leq \binom{d}{nk}.
 \label{eq:aggregate-ball}
\end{align}

The preceding count is the statistical part of the argument: a successful answer restricts the unknown aggregate to a small set. Let $\mathcal S$ be the event of successful reconstruction. Its indicator is determined by $(U,M_1,\ldots,M_n,R)$ and has entropy at most one bit. On success, the remaining uncertainty is bounded by the logarithm of \eqref{eq:aggregate-ball}; on failure, it is bounded by $\log\binom d{nk}$. Thus,
\begin{align*}
 H(U\mid M_1,\ldots,M_n,R)
 &\leq 1+\Prob(\mathcal S)
       \log\!\left(\sum_{j=0}^{\lfloor nD\rfloor}\binom{nk}{j}\binom{d-nk}{j}\right) + \Prob(\mathcal S^c)\log\binom d{nk}\\
 &\leq 1+(1-\delta)
       \log\!\left(\sum_{j=0}^{\lfloor nD\rfloor}\binom{nk}{j}\binom{d-nk}{j}\right) + \delta\log\binom d{nk}.
\end{align*}
The second inequality uses $\Prob(\mathcal S)\geq1-\delta$ and the last bound in \eqref{eq:aggregate-ball}. Subtracting from $H(U\mid R)$ gives
\begin{align}
 I(U;M_1,\ldots,M_n\mid R) &=H(U\mid R)-H(U\mid M_1,\ldots,M_n,R)\notag\\
 &\ge(1-\delta)\log
       \frac{\binom d{nk}}{\displaystyle\sum_{j=0}^{\lfloor nD\rfloor}\binom{nk}{j}\binom{d-nk}{j}}-1.
 \label{eq:aggregate-information}
\end{align}

We next account for the information in the message lengths. The target lower bound is at most $nk\lceil\log d\rceil$. Any protocol with worst-case cost greater than this already satisfies it. We may therefore restrict attention to protocols of cost at most $nk\lceil\log d\rceil$, a cost attainable by sending all active indices. In particular, their expected total length under the chosen prior is at most this value. First, for any nonnegative integer variable $Z$ with mean $\mu>0$, we have
\begin{align*}
 H(Z) \leq \log(\mu+1)+\mu\log(1+1/\mu) \leq \log(\mu+1)+\log e=\log\left(e(\mu+1)\right).
\end{align*}
For a binary string $M$, there are only $2^\ell$ possible messages of length $\ell$. Hence
\[
 H(M\mid |M|,R)\leq\E|M|.
\]
This inequality accounts for the bits within a string, while $H(|M|\mid R)$ accounts for the observed length. Together they avoid imposing any unassumed prefix condition.
The entropy chain rule, subadditivity, and Jensen's inequality then give
\begin{align*}
 I(U;M_1,\ldots,M_n\mid R)
 &\leq \sum_{i=1}^n\left[H(|M_i|)+\E|M_i|\right]\\
 &\leq \sum_{i=1}^n\E|M_i| +n\log\!\left(e\left(\frac1n\sum_{i=1}^n\E|M_i|+1\right)\right)\\
 &\leq \sum_{i=1}^n\E|M_i|
       +n\log\left(e(k\lceil\log d\rceil+1)\right).
\end{align*}
Since the result above holds for any message $(M_1,\ldots,M_n)$, combining it with \eqref{eq:aggregate-information} and taking the infimum, we then obtain
\begin{align}
 \chet(D,\delta) &\geq \inf \sum_{i=1}^n \E[|M_i|]\notag\\*
 &\geq(1-\delta)\log
       \frac{\binom d{nk}}{\displaystyle\sum_{j=0}^{\lfloor nD\rfloor}\binom{nk}{j}\binom{d-nk}{j}} - 1-n\log\left(e(k\lceil\log d\rceil+1)\right).
 \label{eq:aggregate_exact}
\end{align}

Using
\begin{align*}
 \binom d{nk} &=\prod_{j=0}^{nk-1}\frac{d-j}{nk-j}
 \ge\prod_{j=0}^{nk-1}\frac d{nk}
 =\left(\frac d{nk}\right)^{nk} \qquad\text{and}\\
 \sum_{j=0}^{\lfloor nD\rfloor}\binom{nk}{j}\binom{d-nk}{j}&\le\sum_{j=0}^{\lfloor nD\rfloor}\binom{nk}{j}d^j \leq d^{\lfloor nD\rfloor}\sum_{j=0}^{nk}\binom{nk}{j}
 =2^{nk}d^{\lfloor nD\rfloor},
\end{align*}
we approximate the ratio by
\begin{align*}
 \log\frac{\binom d{nk}}{\displaystyle\sum_{j=0}^{\lfloor nD\rfloor}\binom{nk}{j}\binom{d-nk}{j}} \geq nk\log(d/(nk))-nk-\lfloor nD\rfloor\log d = (nk-\lfloor nD\rfloor)\log d-nk\log(2nk).
\end{align*}
Substitution into the lower bound \eqref{eq:aggregate_exact}, followed by
$-(1-\delta)nk\log(2nk)\ge-nk\log(2nk)$, proves \eqref{eq:finite-het-converse}. 
\end{proof}

\subsection{Proof of Theorem~\ref{thm:main}}
\label{app:proof-homogeneous}
\begin{proof}
For all sufficiently large $d$, the relaxed failure probability in \eqref{eq:hom-finite-characterization} is below one. Lemma~\ref{lem:identification} with $\tau=1/\log d$ gives conditionally i.i.d. counts satisfying
\begin{equation}
 \Prob\!\left(\sum_iJ_i<nk-\lfloor nD\rfloor\right)
 \leq\delta+\frac{n^2(k+D)}{(\lfloor nD\rfloor+1-nD)\log d}.
 \label{eq:hom-count-lower-proof}
\end{equation}
Their common conditional law, together with the existing public seed, is feasible in \eqref{eq:lambda}. Therefore
\[
 \E\sum_iJ_i\geq\Lambda_{n,k,D}\!\left(\delta+
 \frac{n^2(k+D)}{(\lfloor nD\rfloor+1-nD)\log d}\right).
\]
Combining this with \eqref{eq:identified-length}, lower bounding worst-case cost by prior-averaged cost, and taking the infimum over protocols proves \eqref{eq:hom-finite-characterization}.

For completeness, the right-continuity used to pass to the limit follows from compactness. The success--cost pairs of common laws form a compact set because the probability simplex is compact and both coordinates are continuous. Its convex hull is also compact and describes all public mixtures. For any sequence $\delta_j\downarrow\delta$, take optimal pairs with success at least $1-\delta_j$. A convergent subsequence has success at least $1-\delta$, so its limiting cost is at least $\Lambda_{n,k,D}(\delta)$. Monotonicity gives the reverse inequality. Hence $\Lambda_{n,k,D}(\delta_j)\to\Lambda_{n,k,D}(\delta)$, including when $\delta=0$.

Divide \eqref{eq:hom-finite-characterization} by $\log d$ and use this continuity. The covering upper bound \eqref{eq:rce_hom_upper} gives
\[
 \frac{\chom(D,\delta)}{\log d}
 \leq\Lambda_{n,k,D}(\delta)\frac{\lceil\log d\rceil}{\log d}
 \longrightarrow\Lambda_{n,k,D}(\delta).
\]
The matching lower limit proves \eqref{eq:hom-range-characterization}.
\end{proof}

\section{Omitted Proofs in Section~\ref{sec:comparison}}
\label{app:comparison-proofs}

\subsection{Proof of Theorem~\ref{thm:comparison}}
\label{app:proof-comparison}
\begin{proof}
The coefficient bounds follow from \eqref{eq:elementary_bounds} in Theorem~\ref{thm:bounds}, together with
\[
 nk-\lfloor nD\rfloor
 \geq n(k-\lfloor D\rfloor)-(n-1)
 \geq k-\lfloor D\rfloor.
\]
If $\lfloor nD\rfloor=n\lfloor D\rfloor$, the upper and lower bounds on $\Lambda_{n,k,D}(\delta)$ in \eqref{eq:elementary_bounds} coincide. To prove the lower bound, suppose $n\ge2$ and
\[
 \Lambda_{n,k,D}(\delta)=(1-\delta)(nk-\lfloor nD\rfloor).
\]
For an optimal $p_R$ of conditionally i.i.d. covering levels $J_i$, decompose the following difference as
\begin{align*}
 0&=\E\sum_iJ_i-(1-\delta)(nk-\lfloor nD\rfloor)\\
 &=\E\!\left[\left(\sum_iJ_i-(nk-\lfloor nD\rfloor)\right)
   \mathds 1\left\{\sum_iJ_i\geq nk-\lfloor nD\rfloor\right\}\right] + \E\!\left[\sum_iJ_i\,
   \mathds 1\left\{\sum_iJ_i<nk-\lfloor nD\rfloor\right\}\right]\\
 &\quad+(nk-\lfloor nD\rfloor)
 \left[\Prob\!\left(\sum_iJ_i\geq nk-\lfloor nD\rfloor\right)-(1-\delta)\right].
\end{align*}
All three terms are nonnegative, and their sum is zero. Therefore, $\sum_iJ_i$ takes values only in $\{0,nk-\lfloor nD\rfloor\}$, with positive probability on the second value. For a non-erasure public randomization, suppose the common level distribution assigns positive probability to two distinct values $a<b$. By conditional independence, the configurations in which all users choose $a$, exactly one user chooses $b$ and the others choose $a$, or all users choose $b$ all have positive probability. For $n\ge2$, their total levels satisfy
\[
 na<(n-1)a+b<nb.
\]
Thus the total level takes at least three distinct values
with positive probability, contradicting its restriction
to $\{0,nk-\lfloor nD\rfloor\}$. Such a randomization must therefore choose a deterministic integer $j$ with $nj=nk-\lfloor nD\rfloor$. Since
\[
 \lfloor D\rfloor\leq \frac{\lfloor nD\rfloor}{n}\leq D<\lfloor D\rfloor+1,
\]
we must have $\lfloor nD\rfloor=n\lfloor D\rfloor$ by the integer constraint of $j$. 
\end{proof}

\subsection{Proof of Theorem~\ref{thm:homogeneity-regimes}}
\label{app:proof-homogeneity-regimes}
\begin{proof}
For fixed $n,k,D,\delta$, the homogeneous characterization
\eqref{eq:hom-range-characterization} gives
\[
 \lim_{d\to\infty}\frac{\chom(D,\delta)}{\log d}
 =\Lambda_{n,k,D}(\delta).
\]
The heterogeneous lower bound \eqref{eq:finite-het-converse} and the
heterogeneous randomized covering construction similarly give
\[
 \lim_{d\to\infty}\frac{\chet(D,\delta)}{\log d}
 =(1-\delta)(nk-\lfloor nD\rfloor).
\]
Since $0\le\delta<1$ and $nk-\lfloor nD\rfloor\ge1$, the latter
coefficient is positive. Dividing the two limits establishes the
asymptotic ratio in \eqref{eq:homogeneity-fixed-parameter-ratio}.

We first prove \eqref{eq:ratio_bound_one}. For fixed $0<\delta<1$ and
$nk-\lfloor nD\rfloor\ge\max\{2,1/\delta\}$, the elementary lower
bound and the concentration upper bound in Theorem~\ref{thm:bounds}
give
\begin{align*}
 (1-\delta)(nk-\lfloor nD\rfloor)
 \le\Lambda_{n,k,D}(\delta) \leq (1-\delta)(nk-\lfloor nD\rfloor) \left[1+O\!\left(\sqrt{\frac{\ln(nk-\lfloor nD\rfloor)}{nk-\lfloor nD\rfloor}}\right)\right].
\end{align*}
In particular, the growth of $nk-\lfloor nD\rfloor$ implies
\[
 \Lambda_{n,k,D}(\delta)
 =(1-\delta)(nk-\lfloor nD\rfloor)(1+o(1)).
\]
The homogeneous randomized covering upper bound \eqref{eq:rce_hom_upper} implies that $\chet(D,\delta)\le\chom(D,\delta) \leq \Lambda_{n,k,D}(\delta)\lceil\log d\rceil$.
Combining this with \eqref{eq:finite-het-converse} gives 
\begin{align*}
 1-&\frac{nk\log(2nk)+1+n\log\bigl(e(k\lceil\log d\rceil+1)\bigr)}
 {(1-\delta)(nk-\lfloor nD\rfloor)\log d}\\
 &\leq \frac{\chet(D,\delta)}
 {(1-\delta)(nk-\lfloor nD\rfloor)\log d}\\
 &\leq \frac{\chom(D,\delta)}
 {(1-\delta)(nk-\lfloor nD\rfloor)\log d} \leq \left[1+O\!\left(\sqrt{\frac{\ln(nk-\lfloor nD\rfloor)}
 {nk-\lfloor nD\rfloor}}\right)\right]
 \frac{\lceil\log d\rceil}{\log d}.
\end{align*}
The growth and remainder conditions in the theorem, together with
$\lceil\log d\rceil/\log d\to1$, make both endpoints tend to one.
This proves \eqref{eq:ratio_bound_one}.

Next, suppose \eqref{eq:ratio-n-distortion} holds. Then $nk-1\leq nD<nk$ and $nk-\lfloor nD\rfloor=1$.
The feasibility condition in \eqref{eq:lambda} requires at least one user to choose a positive covering level. Replacing each individual $J_i$ by $\mathds 1\{J_i\geq1\}$ preserves this event and cannot increase the cost. Thus each $J_i$ may be restricted to $\{0,1\}$, with conditional probabilities $p_{R,0}$ and $p_{R,1}$, respectively; the sum may still take any value in $\{0,\ldots,n\}$. Since the users' $J_1,\ldots,J_n$ are independent and identically distributed given $R$, the probability that all users choose zero is $\E_R[(1-p_{R,1})^n]$.
Hence, the $(D,\delta)$-accuracy constraint under \eqref{eq:ratio-n-distortion} becomes \[
\mathbb E_R[(1-p_{R,1})^n]\le\delta.
\]
By Jensen's inequality,
\[
 (1-\E_R[p_{R,1}])^n
 \le\E_R\bigl[(1-p_{R,1})^n\bigr]\le\delta,
\]
Every feasible Bernoulli-level policy therefore has expected total level $n\E_R[p_{R,1}]\geq n(1-\delta^{1/n})$. Taking the minimum gives $\Lambda_{n,k,D}(\delta)\geq n(1-\delta^{1/n})$.

To show that this lower bound is achievable, consider the case in which every user independently chooses covering level $J_i=1$ with probability $p_{R,1}=p$ and $J_i=0$ with probability $p_{R,0}=1-p$. The covering-level constraint fails exactly when $J_i=0$ for all $i$. By independence, this failure probability is \((1-p)^n\). Choosing \(p=1-\delta^{1/n}\) makes this probability exactly $\delta$. Moreover, each user's expected level is $\E[J_i]=p$, giving an expected total covering level of $np=n(1-\delta^{1/n})$, attaining the lower bound established above and proving
\[
\Lambda_{n,k,D}(\delta)=n(1-\delta^{1/n}).
\]
Combining this identity with the fixed-parameter communication
characterizations proved above gives
\[
 \lim_{d\to\infty}\frac{\chom(D,\delta)}{\chet(D,\delta)}
 =\frac{n(1-\delta^{1/n})}{1-\delta}.
\]
Finally, for fixed $n$, letting $\delta\downarrow0$ gives
\[
 \lim_{\delta\downarrow0}\frac{n(1-\delta^{1/n})}{1-\delta}=n.
\]

\end{proof}

\section{Omitted Proofs in Section~\ref{sec:computation}}
\label{app:computation-proofs}

\subsection{Proof of Lemma~\ref{lem:two-public-modes}}
\label{app:proof-two-modes}
\begin{proof}
First, replacing each covering level $J_i$ by $J_i'=\min\{J_i,nk-\lfloor nD\rfloor\}$ gives \[
\mathds 1\left\{\sum_{i=1}^n J_i' \geq nk-\lfloor nD\rfloor\right\} = \mathds 1\left\{\sum_{i=1}^n J_i \geq nk-\lfloor nD\rfloor\right\},
\]
while
\[
\E\left[\sum_{i=1}^n J_i'\right] \leq \E\left[\sum_{i=1}^n J_i\right].
\]
Hence, we may restrict all covering levels $J_i$ to $\{0,\ldots,\min\{k,nk-\lfloor nD\rfloor\}\}$.

For any distribution $p$ on these levels, define \[
\mathcal K=\operatorname{conv}\left\{\left(\Pr_p\left(\sum_{i=1}^n J_i \geq nk-\lfloor nD\rfloor\right), n\sum_{j=0}^k jp_j\right):p\in\Delta_k\right\},
\]
where $\Delta_k$ denotes the probability simplex restricted
to the admissible covering levels.
Both coordinates are continuous in $p$, so
$\mathcal K$ is compact. By \eqref{eq:lambda},
\[
\Lambda_{n,k,D}(\delta)
=
\min\{y:(x,y)\in\mathcal K,\ x\ge1-\delta\}.
\]

If $\delta=0$, almost every public randomization must succeed with probability one.
Let $a$ be the smallest level with positive probability in such a mode.
Conditional independence gives positive probability to all users choosing $a$,
so $na\geq nk-\lfloor nD\rfloor$. Consequently,
\[
 a\geq\left\lceil k-\frac{\lfloor nD\rfloor}{n}\right\rceil
 =k-\lfloor D\rfloor.
\]
Choosing this level deterministically is feasible and minimizes the cost.
Thus $\Lambda_{n,k,D}(0)=n(k-\lfloor D\rfloor)$, and a single public randomization suffices.

Suppose $0<\delta<1$. Since $(0,0)\in\mathcal K$,
any feasible $(x,y)$ with $x>1-\delta$ can be
replaced by $\frac{1-\delta}{x}(x,y)\in\mathcal K$, which has success probability $1-\delta$ and no larger cost. Thus, $(1-\delta,\Lambda_{n,k,D}(\delta)) \in\partial\mathcal K$.
A supporting line at this point intersects $\mathcal K$ in a point or a line segment. Since the extreme points of $\mathcal K$ belong to the original compact set of success--cost pairs, there exist two distributions $p_r,p_{r'}$ and
$0\le\theta\le1$ such that
\[
1-\delta =\theta\Pr_{p_r}\left( \sum_{i=1}^nJ_i\geq nk-\lfloor nD\rfloor \right) +(1-\theta)\Pr_{p_{r'}}\left(
\sum_{i=1}^nJ_i\geq nk-\lfloor nD\rfloor \right)
\]
and
\[
\Lambda_{n,k,D}(\delta)=\theta n\sum_{j=0}^k jp_{r,j} +(1-\theta)n\sum_{j=0}^k jp_{r',j}.
\]
Thus, at most two public $p_r$ and $p_{r'}$ suffice.
\end{proof}

\subsection{Proof of Theorem~\ref{thm:eptas}}
\label{app:proof-eptas}
\begin{proof}
We first control the cost of each randomization in an optimal mixture. A small average cost alone would permit an expensive randomization with a very small public probability. This control will justify a finite grid that preserves feasibility and approximates the cost.

We first bound $\E[\sum_i J_i]$ within each public randomization appearing with positive probability in an optimal mixture. For fixed $0<\delta<1$, consider an optimal two-randomization mixture $\theta p_r + (1-\theta)p_{r'}$ from Lemma~\ref{lem:two-public-modes}. The set of attainable success--cost pairs has a supporting line of finite nonnegative slope $\lambda$ at the optimal point. Duality for this supporting line gives
\begin{equation}
 \begin{aligned}
 \Lambda_{n,k,D}(\delta)
 =\lambda(1-\delta)-\max_p\Biggl\{&\lambda\Prob_p\!\left(
 \sum_{i=1}^nJ_i\geq nk-\lfloor nD\rfloor\right)\\
 &-n\sum_{j=0}^k jp_j\Biggr\}.
 \end{aligned}
 \label{eq:eptas-dual}
\end{equation}

To bound $\lambda$, consider a reference distribution with independent $J_i\sim\operatorname{Bin}(k,p)$ and
\[
 p=\min\left\{1,\frac{2(nk-\lfloor nD\rfloor)
      +4\lceil\log(2/\delta)\rceil}{nk}\right\}.
\]
The expected total covering level is at most $2(nk-\lfloor nD\rfloor)+4\lceil\log(2/\delta)\rceil$. By the binomial tail calculation in the proof of Theorem~\ref{thm:bounds}, its success probability is at least $1-\delta/2$. Writing the sufficient condition \eqref{eq:binomial_sufficient_condition} from that proof with failure probability $\delta/2$, we have
\[
 \begin{aligned}
 &nk-\lfloor nD\rfloor+\ln\frac2\delta
 +\sqrt{\left(\ln\frac2\delta\right)^2
              +2(nk-\lfloor nD\rfloor)\ln\frac2\delta}\\
 &\quad\leq nk-\lfloor nD\rfloor+2\ln\frac2\delta
          +2\sqrt{(nk-\lfloor nD\rfloor)\ln\frac2\delta}\\
 &\quad\le2(nk-\lfloor nD\rfloor)+3\ln\frac2\delta\\
 &\quad\le2(nk-\lfloor nD\rfloor)+4\lceil\log(2/\delta)\rceil.
 \end{aligned}
\]
When $p=1$, success is certain and the same conclusion holds. Evaluating the maximum in \eqref{eq:eptas-dual} at this reference distribution gives
\[
 \Lambda_{n,k,D}(\delta)
 \le2(nk-\lfloor nD\rfloor)+4\lceil\log(2/\delta)\rceil
       -\frac{\lambda\delta}{2}.
\]
Since $\Lambda_{n,k,D}(\delta)\ge0$,
\[
 0\le\lambda\le
 \frac{4(nk-\lfloor nD\rfloor)+8\lceil\log(2/\delta)\rceil}{\delta}.
\]
The zero-level policy makes the maximum in \eqref{eq:eptas-dual} nonnegative. Each positive-weight randomization of an optimal mixture attains that maximum, so it satisfies
\begin{equation}
 n\sum_{j=1}^k jp_j
 \le\lambda\Prob_p\!\left(\sum_{i=1}^nJ_i\geq nk-\lfloor nD\rfloor\right)
 \le\frac{4(nk-\lfloor nD\rfloor)+8\lceil\log(2/\delta)\rceil}{\delta}.
 \label{eq:mode-cost-bound}
\end{equation}

We now analyze Algorithm~1, starting with its large-threshold branch. The chosen binomial parameter ensures that the non-erasure randomization $p$ fails with probability at most $\varepsilon\delta/8\leq \min\{\delta,\varepsilon/8\}$. Activating this $p$ with public probability $(1-\delta)/(1-\varepsilon\delta/8)$ therefore gives overall success probability at least $1-\delta$. Using the binomial tail calculation in Theorem~\ref{thm:bounds}, 
\[
\begin{aligned}
\frac{\widehat\Lambda_{n,k,D}(\delta)}{\Lambda_{n,k,D}(\delta)} &\le\frac{(1-\delta)nkp}{(1-\varepsilon\delta/8)(1-\delta)t} =\frac{nkp}{(1-\varepsilon\delta/8)t} \le\frac{t+2L+2\lceil\sqrt{tL}\rceil}{(1-\varepsilon\delta/8)t}\\
&\le\frac{1+2L/t+2\sqrt{L/t}+2/t}{1-\varepsilon\delta/8}\\
&\le\frac{1+\varepsilon^2/32+\varepsilon/4+\varepsilon^2/32}{1-\varepsilon/8}\\
&=\frac{1+\varepsilon/4+\varepsilon^2/16}{1-\varepsilon/8}\\
&\le 1+\varepsilon.
\end{aligned}
\]
This proves both feasibility and the claimed accuracy in the large $nk-\lfloor nD\rfloor$ branch.

It remains to consider the small $t = nk-\lfloor nD\rfloor$ case.
Since $h=\min\{k,t\}\le t$, the number of admissible covering levels is bounded by a function of $1/\varepsilon$ and $1/\delta$.
We distinguish two cases according to the number of users.

\smallskip
\noindent\textit{Case 1: Small $n$.}
Suppose
\[
n\le B+\frac{\varepsilon(1-\delta)t}{2}.
\]
Then $n$ is also bounded by a function of
$1/\varepsilon$ and $1/\delta$.
As specified in Algorithm~1, choose $m=\left\lceil\frac{nh}{\varepsilon(1-\delta)t}\right\rceil$.
Consider an optimal mixture of at most two public randomizations,
as established in Lemma~\ref{lem:two-public-modes}.
For each distribution $p_r$ in this mixture,
round its cumulative probabilities downward to multiples
of $1/m$. Specifically, define
\[
p_{r,j}^{+}
=\frac1m\left(
\left\lfloor m\sum_{a=0}^{j}p_{r,a}\right\rfloor
-\left\lfloor m\sum_{a=0}^{j-1}p_{r,a}\right\rfloor
\right),
\quad 0\le j\le h,
\]
where an empty sum is zero by convention.
Consequently, for every $0\le j<h$, $0\le\sum_{a=0}^{j}p_{r,a}-\sum_{a=0}^{j}p_{r,a}^{+}<\frac1m$, implying that $p_r^{+}$ stochastically dominates $p_r$.
Since the users independently draw their covering levels
conditional on each public randomization, this rounding cannot
decrease the probability that their sum reaches $t$.
Moreover, the tail-sum identity gives
\begin{align*}
n\sum_{j=0}^{h}j(p_{r,j}^{+}-p_{r,j}) =n\sum_{j=0}^{h-1}\left(\sum_{a=0}^{j}p_{r,a}-\sum_{a=0}^{j}p_{r,a}^{+}\right)\le\frac{nh}{m}.
\end{align*}

Applying this rounding to both modes, while preserving
their original mixing probabilities, produces a feasible
mixture whose expected total covering level is at most
$\Lambda_{n,k,D}(\delta)+nh/m$.
Since every rounded distribution belongs to the
enumerated family, Algorithm~1 returns a mixture with
no larger expected cost. Consequently,
\begin{align*}
\frac{\widehat\Lambda_{n,k,D}(\delta)}{\Lambda_{n,k,D}(\delta)} 
\le\frac{\Lambda_{n,k,D}(\delta)+nh/m}{\Lambda_{n,k,D}(\delta)}
=1+\frac{nh}{m\Lambda_{n,k,D}(\delta)}
\le1+\frac{nh}{m(1-\delta)t}
\le1+\varepsilon,
\end{align*}
where the last two inequalities follow from the lower
bound \eqref{eq:elementary_bounds} in Theorem~\ref{thm:bounds} and the choice of $m$.

\smallskip
\noindent\textit{Case 2: Large $n$.}
Now suppose
\[
n>B+\frac{\varepsilon(1-\delta)t}{2}.
\]
In this case, Algorithm~1 uses the scaled-probability
grid with spacing $\rho=\frac{\varepsilon(1-\delta)t}{h(h+1)}$.

Consider again an optimal mixture of at most two public randomizations. For each distribution in this mixture, round $np_{r,j}$ upward to the nearest
multiple of $\rho$ for $1\le j\le h$, i.e.,
\[
p_{r,j}^{+}
=\frac{\rho}{n}
\left\lceil\frac{np_{r,j}}{\rho}\right\rceil,
\qquad 1\le j\le h,
\]
and assign the remaining probability $p_{r,0}^{+}=1-\sum_{j=1}^{h}p_{r,j}^{+}$.

We first verify that this rounded distribution is valid.
By \eqref{eq:mode-cost-bound}, each randomization in the
chosen optimal mixture satisfies $n\sum_{j=1}^{h}jp_{r,j}\le B$.
Since $0\le n(p_{r,j}^{+}-p_{r,j})<\rho$,
we obtain
\begin{align*}
n\sum_{j=1}^{h}p_{r,j}^{+}\le n\sum_{j=1}^{h}p_{r,j}+h\rho
\le B+\frac{\varepsilon(1-\delta)t}{h+1}
\le B+\frac{\varepsilon(1-\delta)t}{2}
<n.
\end{align*}
Therefore $p_{r,0}^{+}\ge0$, and $p_r^{+}$
is a valid probability distribution.

Since $p_{r,j}^{+}\ge p_{r,j}$ for every $j\ge1$,
the rounding moves probability from $j=0$ to
$j>0$. Hence, $p_r^{+}$ stochastically
dominates $p_r$, and the success probability cannot
decrease.
Also, the increase in the expected total covering level
of each randomization is bounded by
\begin{align*}
n\sum_{j=1}^{h}j(p_{r,j}^{+}-p_{r,j})
\le\rho\sum_{j=1}^{h}j
=\frac{\rho h(h+1)}{2}
=\frac{\varepsilon(1-\delta)t}{2}.
\end{align*}
Furthermore, \eqref{eq:mode-cost-bound} implies
$np_{r,j}\le B$ for every $j\ge1$. We then have 
\[
a_j =\left\lceil\frac{np_{r,j}}{\rho}\right\rceil
\le\left\lceil\frac B\rho\right\rceil
\le1+\left\lceil\frac B\rho\right\rceil.
\]
Thus each rounded probability belongs to the
scaled grid enumerated in Algorithm~1.

Applying this rounding to both optimal randomizations gives
a feasible mixture with expected total covering level
at most $\Lambda_{n,k,D}(\delta)+\frac{\varepsilon(1-\delta)t}{2}$,
which also upper bounds the cost of the mixture returned by Algorithm~1. Therefore,
\begin{align*}
\frac{\widehat\Lambda_{n,k,D}(\delta)}{\Lambda_{n,k,D}(\delta)}
\le \frac{\Lambda_{n,k,D}(\delta)+\varepsilon(1-\delta)t/2}{\Lambda_{n,k,D}(\delta)}
=1+\frac{\varepsilon(1-\delta)t}{2\Lambda_{n,k,D}(\delta)}
\le1+\frac{\varepsilon}{2}
\le1+\varepsilon.
\end{align*}

Combining the two cases proves the approximation
guarantee for the small-count branch.

Finally, we bound the arithmetic complexity of the algorithm.
In the small-$t$ branch, both $t$ and $h$ are bounded by
functions of $\varepsilon$ and $\delta$.
When $n\le B+\varepsilon(1-\delta)t/2$, the number of
candidate distributions is
$\binom{m+h}{h}\le(m+1)^h$, where
$m=\lceil nh/[\varepsilon(1-\delta)t]\rceil$.
When $n>B+\varepsilon(1-\delta)t/2$, the number of candidates
is at most
\[
\left(2+\left\lceil\frac{B}{\rho}\right\rceil\right)^h,
\qquad
\frac{B}{\rho}
=O\left(
\frac{\log^2(8/(\varepsilon\delta))}
{\delta(1-\delta)\varepsilon^5}
\right).
\]
Thus, in either case, the number $N$ of candidate distributions
depends only on $\varepsilon$ and $\delta$, with
\[
N\le
\exp\left[
O\left(
\frac{\log(8/(\varepsilon\delta))}{\varepsilon^2}
\log\left(
\frac{\log^2(8/(\varepsilon\delta))}
{\delta(1-\delta)\varepsilon^5}
\right)
\right)
\right].
\]

For each candidate, the success probability can be evaluated
by truncated polynomial exponentiation using
$O(t^2\log(n+1))$ arithmetic operations, and its expected
covering level requires $O(h)$ operations.
The lower convex hull of the resulting success--cost pairs
can be computed in $O(N\log N)$ operations.
Hence, the small-$t$ branch requires
\[
O\left(
N\left[t^2\log(n+1)+h+\log N\right]
\right)
\]
arithmetic operations. The large-$t$ branch requires at most
$O(k)$ operations to construct the binomial distribution,
and explicitly representing the final distribution over
all $k+1$ covering levels also requires at most $O(k)$
operations.

Since $t$, $h$, and $N$ are bounded by functions of
$\varepsilon$ and $\delta$, the polynomial factors in $t$, $h$, and $\log N$ can be absorbed into the exponential bound above. Consequently, the overall arithmetic complexity is \[
O\left(k+f(1/\varepsilon,1/\delta)\log(n+1)\right),
\]
where
\[
f\left(\frac{1}{\varepsilon},\frac{1}{\delta}\right)
=
\exp\left[
O\left(
\frac{\log(8/(\varepsilon\delta))}{\varepsilon^2}
\log\left(
\frac{\log^2(8/(\varepsilon\delta))}
{\delta(1-\delta)\varepsilon^5}
\right)
\right)
\right].
\]
For every fixed $0<\delta<1$, this establishes the
claimed arithmetic complexity.

\end{proof}

\section{Derivation and Numerical Evaluation of the Comparison Bounds}
\label{app:quantitative-comparison}

All baseline constructions admit a common conditional encoder. Public randomness may choose among parameter settings, including a setting in which all users remain silent. We choose these probabilities to minimize expected communication subject to the weighted average of the uniform success bounds being at least $1-\delta$.

\subsection{Coordinate and Client Sampling}

Both active sampling estimators reconstruct a reported entry as $1/p$. Put $q=nk$ for coordinate sampling and $q=n$ for client sampling. For $1/2<p\leq1$, summing local absolute errors gives
\begin{equation}
 \Pr(\|\hat u-u\|_1\leq D)
 \geq \Pr\!\left\{\operatorname{Bin}(q,p)\geq
 \left\lceil\frac{qp(1-D/k)}{2p-1}\right\rceil\right\}.
 \label{eq:sampling-binomial-comparison}
\end{equation}
Indeed, the sum of local errors for inputs with exactly $k$ active entries is distributed as
$k+(k/q)(p^{-1}-2)\operatorname{Bin}(q,p)$.
For smaller supports, adding nonnegative dummy error contributions gives the same upper bound. The expected communication is at most $nkp\lceil\log d\rceil$. 

For the explicit table bounds, Hoeffding's inequality and $0<2-p^{-1}\leq1$ give
\[
 \|\hat u-u\|_1
 \leq 2k(1-p)+k\sqrt{\frac{\ln(2/\delta)}{2q}}
\]
with probability at least $1-\delta/2$. Choose
\[
 p=\min\left\{1,\,
 1-\frac{D}{2k}+\sqrt{\frac{\ln(2/\delta)}{8q}}\right\}.
\]
When $p<1$, use this sampling rule with public probability $(1-\delta)/(1-\delta/2)$ and let all users remain silent otherwise. When $p=1$, exact transmission with probability $1-\delta$ suffices. Comparing these options gives expected communication at most
\begin{equation}
 (1-\delta)nk\lceil\log d\rceil
 \min\left\{1,\,
 \frac{1-D/(2k)+\sqrt{\ln(2/\delta)/(8q)}}{1-\delta/2}\right\}.
 \label{eq:sampling-explicit-comparison}
\end{equation}
Substituting $q=nk$ and $q=n$ proves the two sampling entries. Their coarse order is $O((1-\delta)nk\log d)$ on $D<k$; the displayed terms show the distortion dependence and the different concentration corrections.

\subsection{Hashing and Algebraic Recovery}

Count-Min thresholded at $1$ and partitioned Bloom filters have no false negatives. With $w>k$ buckets in each of $r$ independent rows, a zero coordinate is falsely reported with probability at most $(k/w)^r$. The aggregate $\ell_1$ error equals the total number of local false positives divided by $n$. Failure therefore requires at least $\lfloor nD\rfloor+1$ false positives, and Markov's inequality gives
\begin{equation}
 \Pr(\|\hat u-u\|_1>D)
 \leq \min\left\{1,\,
 \frac{nd}{\lfloor nD\rfloor+1}\left(\frac{k}{w}\right)^r\right\}.
 \label{eq:sketch-error-budget}
\end{equation}
Independence across users or queried coordinates is unnecessary. A counter row has at most $\binom{w+k}{k}$ possible states, and an occupancy row has at most $\sum_{j=0}^k\binom wj$. The respective total communication costs of Count-Min Sketch and Bloom Filters are at most
\[
 nr\left\lceil\log\binom{w+k}{k}\right\rceil,
 \qquad
 nr\left\lceil\log\sum_{j=0}^k\binom wj\right\rceil.
\]
For the table, take
\[
 w=2k,\qquad
 r=\left\lceil\log\frac{2nd}{\delta(\lfloor nD\rfloor+1)}\right\rceil.
\]
The failure bound is then at most $\delta/2$. Each row costs $O(k)$ bits, and
$(1+nD)/2\leq\lfloor nD\rfloor+1\leq1+nD$.
Using the sketch with public probability $(1-\delta)/(1-\delta/2)$ and remaining silent otherwise proves both hashing entries in Table~\ref{tab:prior-work}.

A BCH syndrome uses at most $k\lceil\log(d+1)\rceil$ bits per user~\cite[Theorem 16]{belazzougui2024reconciliation}. Exact decoding with public transmission probability $1-\delta$ gives $(1-\delta)nk\lceil\log(d+1)\rceil$ expected bits.

\section{Lower Bounds Comparisons under Expected-Length Communication}
\label{app:converse-comparison}

We justify Table~\ref{tab:converse-comparison} by adapting the three lower bound methods to expected communication and our accuracy criterion.

\subsection{From Information to Expected Bits}

The geometric distribution maximizes the entropy of a nonnegative integer with a fixed mean. Applying this fact to the message lengths, and then counting the binary strings of each length, gives
\begin{equation}
 \begin{aligned}
 I(X_1,\ldots,X_n;M_1,\ldots,M_n\mid R)
 &\leq C+(C+n)\log(1+C/n)-C\log(C/n).
 \end{aligned}
 \label{eq:comparison-length-entropy}
\end{equation}
Here $C$ is the worst-case expected total communication, and $0\log0=0$. The right-hand side is increasing in $C$, so an information lower bound gives a communication lower bound by solving this scalar inequality. For fixed $n$ and information of order $\log d$, the message-length entropy changes the bound by only $O(\log\log d)$.

\subsection{Aggregate Fano and Transcript Counting}

For $d\geq nk$, choose a uniform $nk$-subset and partition it into $n$ supports of size $k$. By \eqref{eq:aggregate-ball}, every distortion ball of radius $D$, including those centered at real-valued outputs, has prior mass at most
\begin{equation}
 \frac{\displaystyle\sum_{j=0}^{\lfloor nD\rfloor}
              \binom{nk}j\binom{d-nk}j}
      {\binom d{nk}}.
 \label{eq:comparison-ball-information}
\end{equation}
Binomial coefficients outside their usual range are zero. The distance-based Fano inequality~\cite{duchi2013distance,fan-cover} and \eqref{eq:comparison-length-entropy} give
\begin{equation}
 \begin{aligned}
 &C+(C+n)\log(1+C/n)-C\log(C/n)\\
 &\quad\geq\left[
 (1-\delta)\log
 \frac{\binom d{nk}}
      {\displaystyle\sum_{j=0}^{\lfloor nD\rfloor}
          \binom{nk}j\binom{d-nk}j}
 +\delta\log\delta+(1-\delta)\log(1-\delta)
 \right]_+.
 \end{aligned}
 \label{eq:comparison-fano-finite}
\end{equation}

For counting~\cite{fan-cover}, the number of message tuples with total length at most a nonnegative integer $\ell$ is
$\sum_{s=0}^{\ell}2^s\binom{s+n-1}{n-1}$.
Conditional on public randomness, each tuple can account for successful prior mass no larger than \eqref{eq:comparison-ball-information}. Bounding the probability of success with total length at most $\ell$, then summing the complementary tail probabilities, gives
\begin{equation}
 C\geq\sum_{\ell=0}^{\infty}\left[
 1-\delta-
 \frac{\displaystyle\sum_{j=0}^{\lfloor nD\rfloor}
                    \binom{nk}j\binom{d-nk}j}
      {\binom d{nk}}
 \sum_{s=0}^{\ell}2^s\binom{s+n-1}{n-1}
 \right]_+.
 \label{eq:comparison-count-finite}
\end{equation}
For fixed $n,k,D,\delta$, the negative logarithm of \eqref{eq:comparison-ball-information} is
$(nk-\lfloor nD\rfloor)\log d+O(1)$.
The logarithm of the tuple count is $\ell+O(\log(\ell+1))$. Both arguments therefore give
\[
 C\geq(1-\delta)(nk-\lfloor nD\rfloor)\log d-O(\log\log d).
\]
Neither uses the common encoder constraint, so both give the same bound for the two models.

\subsection{Posterior Variance}

Following the block prior in~\cite{huang2019optimal}, choose $k$ disjoint groups of $\lfloor d/k\rfloor\geq2$ coordinates. Each user independently chooses one active coordinate in each group. Euclidean projection of the estimate onto $\{v\in[0,1]^d:\|v\|_1\leq k\}$ does not increase its squared error. The projected squared error is at most $D^2$ on success and at most $2k$ on every outcome. Its mean is therefore at most
\[
 (1-\delta)D^2+2k\delta.
\]
The conditional variance decomposition in~\cite[Lemma 3.4]{huang2019optimal} bounds the expected sum of local posterior variances by
$n^2((1-\delta)D^2+2k\delta)$.
For a categorical variable, its minimum decoding error is at most the sum of its indicator variances. Thus the same expression bounds the sum of the block-decoding error probabilities.

The standard decoding-error entropy bound limits each block's conditional entropy to one bit plus its decoding error probability times $\log(\lfloor d/k\rfloor-1)$. Summing over the $nk$ blocks gives
\[
 \begin{aligned}
 I(X_1,\ldots,X_n;M_1,\ldots,M_n\mid R)
 \geq{}&nk\log\lfloor d/k\rfloor-nk\\
 &-n^2((1-\delta)D^2+2k\delta)
       \log(\lfloor d/k\rfloor-1).
 \end{aligned}
\]
Information is nonnegative. Combining these bounds with \eqref{eq:comparison-length-entropy}, for fixed $n,k,D,\delta$, gives
\begin{equation}
 C\geq\big[nk-n^2((1-\delta)D^2+2k\delta)\big]_+
       \log d-O(\log\log d).
 \label{eq:comparison-variance-finite}
\end{equation}
This proves the coefficient in Table~\ref{tab:converse-comparison}. Both this argument and our lower bounds use local posterior information. The squared-loss conversion replaces the success-event requirement by an average variance bound. Our identification lemma instead controls how many true support coordinates the messages must identify on successful outcomes and preserves their common conditional law in the homogeneous model.

\bibliographystyle{IEEEtran}
\bibliography{ref}

@inproceedings{huang2019optimal,
  author    = {Huang, Zengfeng and Huang, Ziyue and Wang, Yilei and Yi, Ke},
  title     = {Optimal Sparsity-Sensitive Bounds for Distributed Mean Estimation},
  booktitle = {Advances in Neural Information Processing Systems},
  volume    = {32},
  year      = {2019}
}

@inproceedings{jhunjhunwala2021spatial,
  author    = {Jhunjhunwala, Divyansh and Mallick, Ankur and Gadhikar, Advait and Kadhe, Swanand and Joshi, Gauri},
  title     = {Leveraging Spatial and Temporal Correlations in Sparsified Mean Estimation},
  booktitle = {Advances in Neural Information Processing Systems},
  volume    = {34},
  pages     = {14280--14292},
  year      = {2021},
  url       = {https://neurips.cc}
}

@inproceedings{vargaftik2022eden,
  author    = {Vargaftik, Shay and Ben Basat, Ran and Portnoy, Amit and Mendelson, Gal and Ben-Itzhak, Yaniv and Mitzenmacher, Michael},
  title     = {{EDEN}: Communication-Efficient and Robust Distributed Mean Estimation for Federated Learning},
  booktitle = {Proceedings of the 39th International Conference on Machine Learning},
  series    = {Proceedings of Machine Learning Research},
  volume    = {162},
  pages     = {21984--22012},
  year      = {2022},
  publisher = {PMLR},
  url       = {https://mlr.press}
}

@article{cormode2005countmin,
  author     = {Cormode, Graham and Muthukrishnan, S.},
  title      = {An improved data stream summary: the count-min sketch and its applications},
  journal    = {Journal of Algorithms},
  volume     = {55},
  number     = {1},
  pages      = {58--75},
  year       = {2005},
  publisher  = {Elsevier},
  doi        = {10.1016/j.jalgor.2003.12.001},
  url        = {https://doi.org}
}

@inproceedings{mcmahan2017federated,
  author    = {McMahan, Brendan and Moore, Eider and Ramage, Daniel and Hampson, Seth and {\relax Ag{\"u}era y Arcas}, Blaise},
  title     = {Communication-Efficient Learning of Deep Networks from Decentralized Data},
  booktitle = {Proceedings of the 20th International Conference on Artificial Intelligence and Statistics},
  series    = {Proceedings of Machine Learning Research},
  volume    = {54},
  pages     = {1273--1282},
  year      = {2017},
  publisher = {PMLR},
  url       = {https://proceedings.mlr.press/v54/mcmahan17a.html}
}

@inproceedings{suresh2017mean,
  title={Distributed Mean Estimation with Limited Communication},
  author={Suresh, Ananda Theertha and Yu, Felix X. and Kumar, Sanjiv and McMahan, H. Brendan},
  booktitle={Proceedings of the 34th International Conference on Machine Learning (ICML)},
  pages={3329--3337},
  year={2017},
  series={Proceedings of Machine Learning Research},
  volume={70},
  publisher={PMLR}
}

@article{konecny2018mean,
  author    = {Kone{\v{c}}n{\'{y}}, Jakub and Richt{\'{a}}rik, Peter},
  title     = {Randomized Distributed Mean Estimation: Accuracy vs Communication},
  journal   = {Frontiers in Applied Mathematics and Statistics},
  volume    = {4},
  pages     = {62},
  year      = {2018},
  publisher = {Frontiers},
  doi       = {10.3389/fams.2018.00062},
  url       = {https://doi.org}
}

@misc{fan-cover,
  author = {Yijun Fan and Fangwei Ye and Raymond W. Yeung},
  title = {Lossy Compression for Sparse Aggregation},
  year = {2026},
  howpublished = {arXiv:2606.30425},
  eprint = {2606.30425},
  archivePrefix = {arXiv},
  primaryClass = {cs.IT},
  url = {https://arxiv.org/abs/2606.30425}
}

@inproceedings{acharya2019communication,
  title={Communication Complexity in Locally Private Distribution Estimation and Heavy Hitters},
  author={Acharya, Jayadev and Sun, Ziteng},
  booktitle={Proceedings of the 36th International Conference on Machine Learning (ICML)},
  pages={51--60},
  year={2019},
  series={Proceedings of Machine Learning Research},
  volume={97},
  publisher={PMLR}
}

@article{acharya2020inference,
  author = {Jayadev Acharya and Cl{\'e}ment L. Canonne and Himanshu Tyagi},
  title = {Inference under Information Constraints {II}: Communication Constraints and Shared Randomness},
  journal = {IEEE Transactions on Information Theory},
  volume = {66},
  number = {12},
  pages = {7856--7877},
  year = {2020},
  doi = {10.1109/TIT.2020.3028439}
}

@inproceedings{chen2023security,
  title={The Communication Cost of Security and Privacy in Federated Frequency Estimation},
  author={Chen, Wei-Ning and {\"O}zg{\"u}r, Ayfer and Cormode, Graham and Bharadwaj, Akash},
  booktitle={Proceedings of the 26th International Conference on Artificial Intelligence and Statistics (AISTATS)},
  pages={4247--4274},
  year={2023},
  series={Proceedings of Machine Learning Research},
  volume={206},
  publisher={PMLR}
}

@article{kostina2015variable,
  author = {Victoria Kostina and Yury Polyanskiy and Sergio Verd{\'u}},
  title = {Variable-Length Compression Allowing Errors},
  journal = {IEEE Transactions on Information Theory},
  volume = {61},
  number = {8},
  pages = {4316--4330},
  year = {2015},
  doi = {10.1109/TIT.2015.2438831}
}

@inproceedings{braverman2016distributed,
  author = {Mark Braverman and Ankit Garg and Tengyu Ma and Huy L. Nguyen and David P. Woodruff},
  title = {Communication Lower Bounds for Statistical Estimation Problems via a Distributed Data Processing Inequality},
  booktitle = {Proceedings of the 48th Annual ACM Symposium on Theory of Computing},
  pages = {1011--1020},
  year = {2016},
  publisher = {ACM},
  doi = {10.1145/2897518.2897582}
}

@inproceedings{suresh2022correlated,
  title={Correlated Quantization for Distributed Mean Estimation and Optimization},
  author={Suresh, Ananda Theertha and Sun, Ziteng and Ro, Jae Hun and Yu, Felix},
  booktitle={Proceedings of the 39th International Conference on Machine Learning (ICML)},
  pages={20758--20771},
  year={2022},
  series={Proceedings of Machine Learning Research},
  volume={162},
  publisher={PMLR}
}

@misc{duchi2013distance,
  author = {John C. Duchi and Martin J. Wainwright},
  title = {Distance-Based and Continuum {Fano} Inequalities with Applications to Statistical Estimation},
  year = {2013},
  howpublished = {arXiv:1311.2669},
  eprint = {1311.2669},
  archivePrefix = {arXiv},
  primaryClass = {cs.IT}
}

@book{boucheron2013concentration,
  title={Concentration Inequalities: A Nonasymptotic Theory of Independence},
  author={Boucheron, St{\'e}phane and Lugosi, G{\'a}bor and Massart, Pascal},
  year={2013},
  publisher={Oxford University Press},
  address={Oxford, UK},
  isbn={978-0-19-953525-5}
}

@article{bloom1970space,
  author = {Burton H. Bloom},
  title = {Space/Time Trade-offs in Hash Coding with Allowable Errors},
  journal = {Communications of the ACM},
  volume = {13},
  number = {7},
  pages = {422--426},
  year = {1970},
  doi = {10.1145/362686.362692}
}

@article{mitzenmacher2002compressed,
  title={Compressed bloom filters},
  author={Mitzenmacher, Michael},
  journal={IEEE/ACM Transactions on Networking},
  volume={10},
  number={5},
  pages={604--612},
  year={2002},
  publisher={IEEE}
}

@inproceedings{belazzougui2024reconciliation,
  author    = {Belazzougui, Djamal and Kucherov, Gregory and Walzer, Stefan},
  title     = {Better Space-Time-Robustness Trade-Offs for Set Reconciliation},
  booktitle = {51st International Colloquium on Automata, Languages, and Programming (ICALP 2024)},
  series    = {Leibniz International Proceedings in Informatics (LIPIcs)},
  volume    = {297},
  pages     = {20:1--20:19},
  year      = {2024},
  publisher = {Schloss Dagstuhl -- Leibniz-Zentrum f{\"u}r Informatik},
  doi       = {10.4230/LIPIcs.ICALP.2024.20},
  url       = {https://doi.org/10.4230/LIPIcs.ICALP.2024.20}
}

@inproceedings{esfandiari2024coordinator,
  author    = {Esfandiari, Hossein and Kacham, Praneeth and Mirrokni, Vahab and Woodruff, David P. and Zhong, Peilin},
  title     = {Optimal Communication Bounds for Classic Functions in the Coordinator Model and Beyond},
  booktitle = {Proceedings of the 56th Annual ACM Symposium on Theory of Computing (STOC 2024)},
  pages     = {1911--1922},
  year      = {2024},
  publisher = {ACM},
  doi       = {10.1145/3618260.3649742},
  url       = {https://doi.org/10.1145/3618260.3649742}
}

@inproceedings{feng2025heavyhitters,
  author    = {Feng, Shiyuan and Swartworth, William and Woodruff, David P.},
  title     = {Tight Bounds for Heavy-Hitters and Moment Estimation in the Sliding Window Model},
  booktitle = {52nd International Colloquium on Automata, Languages, and Programming (ICALP 2025)},
  series    = {Leibniz International Proceedings in Informatics (LIPIcs)},
  volume    = {334},
  pages     = {75:1--75:19},
  year      = {2025},
  publisher = {Schloss Dagstuhl -- Leibniz-Zentrum f{\"u}r Informatik},
  doi       = {10.4230/LIPIcs.ICALP.2025.75},
  url       = {https://doi.org}
}

@article{bhattacharyya2018heavyhitters,
  author     = {Bhattacharyya, Sayan and Dey, Palash and Woodruff, David P.},
  title      = {An Optimal Algorithm for $\ell_1$-Heavy Hitters in Insertion Streams and Related Problems},
  journal    = {ACM Transactions on Database Systems (TODS)},
  volume     = {43},
  number     = {2},
  pages      = {9:1--9:26},
  year       = {2018},
  publisher  = {ACM New York, NY, USA},
  doi        = {10.1145/3264427},
  url        = {https://doi.org}
}

@inproceedings{karnin2016quantiles,
  author    = {Karnin, Zohar and Lang, Kevin and Liberty, Edo},
  title     = {Optimal Quantile Approximation in Streams},
  booktitle = {2016 IEEE 57th Annual Symposium on Foundations of Computer Science (FOCS)},
  pages     = {420--429},
  year      = {2016},
  publisher = {IEEE},
  doi       = {10.1109/FOCS.2016.52},
  url       = {https://doi.org}
}

@inproceedings{braverman2016countsketch,
  author    = {Braverman, Vladimir and Chestnut, Stephen R. and Ivkin, Nikita and Woodruff, David P.},
  title     = {Beating {CountSketch} for Heavy Hitters in Insertion Streams},
  booktitle = {Proceedings of the Forty-Eighth Annual ACM Symposium on Theory of Computing (STOC 2016)},
  pages     = {740--753},
  year      = {2016},
  publisher = {ACM},
  doi       = {10.1145/2897518.2897558},
  url       = {https://doi.org}
}

@inproceedings{liu2020succinctfilters,
  author    = {Liu, Mingmou and Yin, Yitong and Yu, Huacheng},
  title     = {Succinct Filters for Sets of Unknown Sizes},
  booktitle = {47th International Colloquium on Automata, Languages, and Programming (ICALP 2020)},
  series    = {Leibniz International Proceedings in Informatics (LIPIcs)},
  volume    = {168},
  pages     = {79:1--79:14},
  year      = {2020},
  publisher = {Schloss Dagstuhl -- Leibniz-Zentrum f{\"u}r Informatik},
  doi       = {10.4230/LIPIcs.ICALP.2020.79},
  url       = {https://doi.org}
}
\end{document}